\documentclass{PoS}

\usepackage{amsmath,amstext,amsfonts,amsbsy,amssymb,amscd,bbm,epsfig}
\usepackage{color,lscape,multirow}
\usepackage{CJKutf8}

\newcommand{\ba}{\begin{array}}
\newcommand{\ea}{\end{array}}

\newcommand{\req}[1]{Eq.~(\ref{#1})}
\newcommand{\res}[1]{Section~\ref{#1}}

\newcommand{\refig}[1]{Fig.~\ref{#1}}
\newcommand{\ret}[1]{Table~\ref{#1}}

\newcommand{\dif}{{\rm d}}

\newcommand{\Dslash}{\relax{\kern+.25em / \kern-.70em D}}
\newcommand{\GF}{G_{\rm\scriptscriptstyle F}}

\newcommand{\fm}{{\rm fm}}
\newcommand{\MeV}{{\rm MeV}}

\newcommand{\Real}{\relax{\mathsf{\Gamma\kern-.35em R}}}
\newcommand{\Int}{\relax{\mathsf{Z\kern-.40em Z}}}

\newcommand{\quart}{{\scriptstyle{{1\over 4}}}}

\newcommand{\NF}{N_\mathrm{\scriptstyle f}}

\newcommand{\gbar}{\kern1pt\overline{\kern-1pt g\kern-0pt}\kern1pt}
\newcommand{\mbar}{\kern2pt\overline{\kern-1pt m\kern-1pt}\kern1pt}
\newcommand{\obar}[1]{\kern3pt\overline{\kern-2pt #1\kern-0pt}\kern1pt}

\newcommand{\lQCD}{\Lambda_{\rm\scriptscriptstyle QCD}}

\newcommand{\orgi}[1]{\hat #1}

\newcommand{\Oa}{\mbox{O}(a)}

\newcommand{\abar}{\kern1pt\overline{\kern-1pt a\kern-0.5pt}\kern1pt}

\newcommand{\cB}{{\cal B}}

\newcommand{\cF}{{\cal F}}
\newcommand{\cG}{{\cal G}}

\newcommand{\cO}{{\cal O}}

\newcommand{\vp}{\mathbf{p}}

\newdimen\arrayruleHwidth
\makeatletter
\def\Hline{\noalign{\ifnum0=`}\fi\hrule \@height \arrayruleHwidth
  \futurelet \@tempa\@xhline}
\makeatother
\let\OLDthebibliography\thebibliography
\renewcommand\thebibliography[1]{
  \OLDthebibliography{#1}
  \setlength{\parskip}{0.5pt}
  \setlength{\itemsep}{0.5pt plus 0.3ex}
}

\title{Progress and prospects for heavy flavour physics on the lattice}

\ShortTitle{Progress and prospects for heavy flavour physics on the lattice}

\author{\speaker{Carlos Pena}\\%
        Department of Theoretical Physics and IFT UAM-CSIC\\
        Universidad Aut\'onoma de Madrid\\
        E-28049 Madrid, Spain\\
        E-mail: \email{carlos.pena@uam.es}}

\abstract{I review recent progress in lattice computations relevant for $B$- and charm physics,
focusing on decay and mixing amplitudes with a direct impact on CKM analysis.
Emphasis is put on the interplay with the upcoming new generation of experimental results,
and the subsequent challenges for lattice computations in the heavy quark sector.}

\FullConference{The 33rd International Symposium on Lattice Field Theory\\
		 14 -18 July  2015\\
		 Kobe International Conference Center, Kobe, Japan}

\begin{document}\begin{CJK}{UTF8}{min}

\section{Introduction}
\label{sec:intro}

\noindent
Flavour physics has long been one of the most powerful tools in the search
for new physics, having led to key milestones in the construction of the Standard Model (SM).
Nowadays, charm and bottom physics at the intensity frontier
are pivotal to efforts to probe the limits of the Standard Model (SM).

Lattice QCD (LQCD) is the only known first-principles approach to the
computation of long-distance strong-interaction contributions to flavour-changing
processes in the quark sector. The maturity reached by simulation techniques in the last decade, 
which has resulted in a dramatic improvement of LQCD predictions for light quark
physics, has more recently had a similar impact on the heavy quark sector:
as will be discussed below, many decay and mixing processes that play a key
role in the determination of Cabibbo-Kobayashi-Maskawa (CKM) matrix elements,
as well as in unitarity triangle analysis of CKM consistency, can be now determined
to very good precision. This in turn allows to fully exploit the large amount
of experimental information produced by the BaBar and Belle B-factories.
Yet, the new generation of results produced by such experiments as LHCb or BESIII,
and, above all, the dramatic improvement in precision expected from the Belle II
experiment~\cite{Iijima:latt15}, will pose a significant challenge to LQCD computations in the
near future.

This review will focus on the computation of SM weak decay and mixing amplitudes
of heavy mesons. A number of important topics (such as heavy quark masses, which
require the discussion of a wholly different theoretical toolset, or amplitudes relevant
for new physics models) will be sacrificed to brevity. New results for leptonic decays
and $B^0$--$\bar B^0$ mixing will be summarised rapidly; more emphasis will be made
on semileptonic decays, for which the last two years have witnessed a comparatively
much more significant progress. The discussion will be mostly structured around
the update of the Flavour Lattice Averaging Group (FLAG) review~\cite{FLAGpre}, the third edition
of which is now finalised~\cite{FLAG3}. Detailed information about the new results presented
at the conference can be found in the relevant proceedings contributions
\cite{latt15global,DeTar:latt15,Primer:latt15,Lami:latt15,Suzuki:latt15}.


\section{Reach of and formalisms for heavy quark physics on the lattice}
\label{sec:reachform}

\subsection{Heavy quark scales in lattice simulations}

\begin{figure}[t!]
\begin{center}
\begin{minipage}[t]{0.590\textwidth}
\vspace{0pt}
\includegraphics[width=\textwidth]{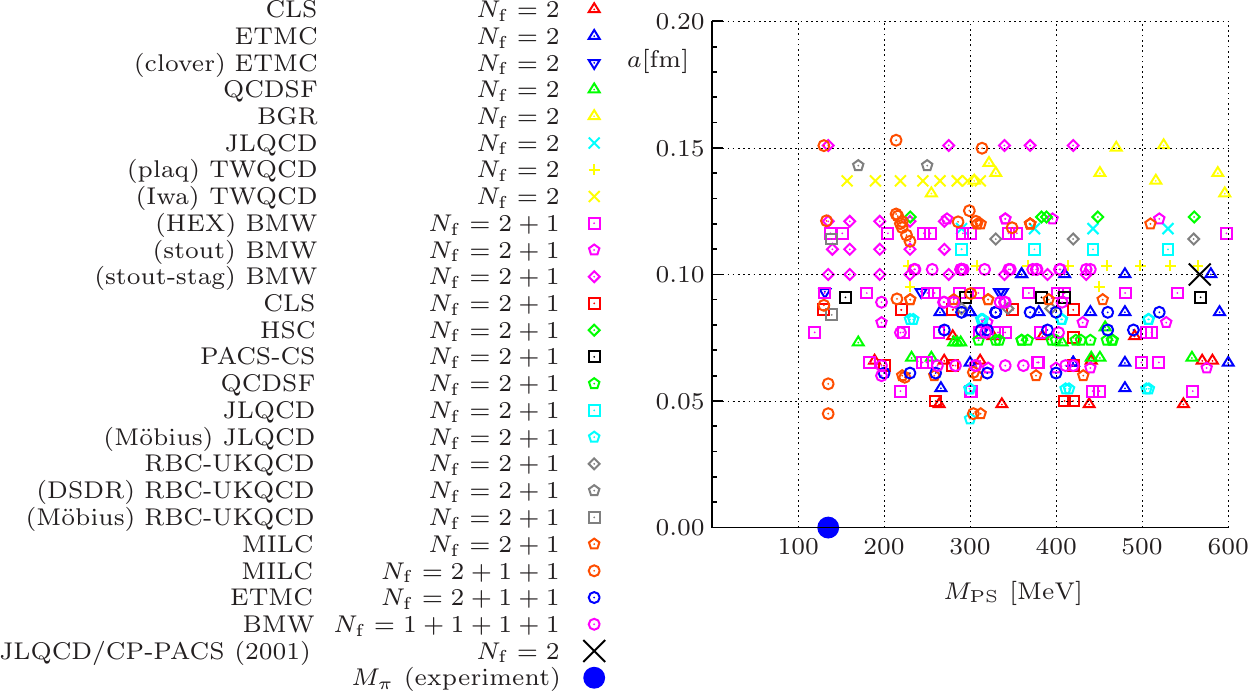}
\end{minipage}
\hspace{10mm}
\begin{minipage}[t]{0.325\textwidth}
\vspace{4pt}
\includegraphics[width=\textwidth]{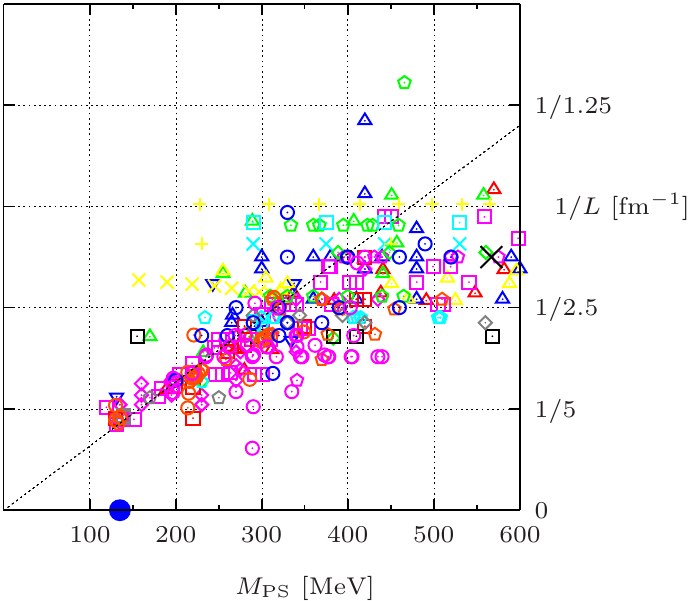}
\end{minipage}
\end{center}
\begin{center}
\begin{minipage}[t]{0.590\textwidth}
\vspace{-6pt}
\includegraphics[width=\textwidth]{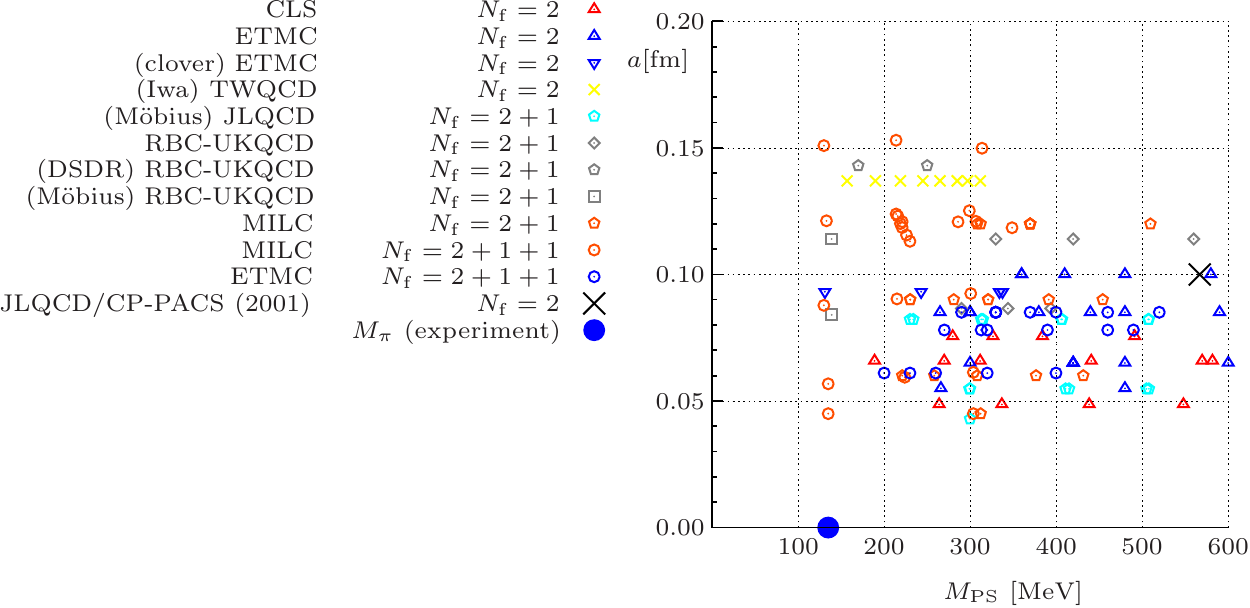}
\end{minipage}
\hspace{10mm}
\begin{minipage}[t]{0.325\textwidth}
\vspace{-2pt}
\includegraphics[width=\textwidth]{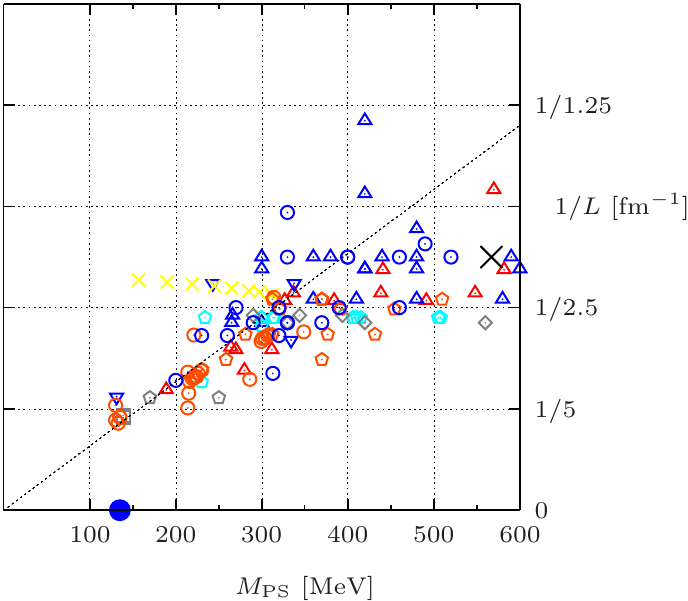}
\end{minipage}
\end{center}
\vspace{-6mm}
\caption{Top: LQCD simulation landscape as of Summer 2015. Ensembles are
shown in the plane spanned by the sea pion mass $M_{\rm PS}$ and the lattice spacing $a$ (left),
and in the plane spanned by $M_{\rm PS}$ and the spatial box size $L$ (right;
the straight line corresponds to $M_{\rm PS}L=4$).
Bottom: same as above, now showing only the ensembles used in computations that will be discussed
in this review. Figures courtesy of G.~Herdo\'{\i}za.}
\label{fig:simulation_landscape}
\end{figure}

\noindent
Any LQCD simulation involves an ultraviolet cutoff, given by the (inverse of the)
lattice spacing $a$, and an infrared cutoff, given by the inverse of the spatial box size $L$.
In order to bring cutoff effects under control, and allow for well-controlled
continuum and infinite-volume limit extrapolations, all physical mass scales $M$ in the problem
addressed need to be far away from the cutoffs, $L^{-1} \ll M \ll a^{-1}$.
Apart from the intrinsic strong interaction scale $\lQCD$, this applies to
the values of hadron masses for all the flavours involved in the computation.
Infrared cutoff effects most often take the
form of finite-volume corrections to correlation functions, which for large
enough volumes ($m_\pi L \gtrsim 4$ being a reasonable rule of thumb) behave
$\sim e^{-m_\pi L}$; this implies that computations at the physical values
of light quark masses must take place in boxes of size $L \gtrsim 6~\fm$,
and pions twice heavier than their physical mass still require $L \gtrsim 3~\fm$.
Ultraviolet cutoff effects, on the other hand, are generally expected to be power-like,
based on Symanzik effective theory. Assuming that all $\Oa$ effects are removed,
either by the properties of the regularisation, or by the inclusion of appropriate
counterterms to the lattice action and composite operators, the leading cutoff effects
driven by quark masses will be $\sim (am_q)^2$. This means that lattice spacings
$a \lesssim 0.1~\fm$ are (naively) enough to keep effects related to the light $u,d,s$ quarks at the
few percent level, but significantly smaller lattice spacings are needed for
charm- and $B$-physics: for instance, $(am_q)^2 \lesssim 0.2$ implies $a \lesssim 0.07~\fm$ for $q=c$,
and $a \lesssim 0.02~\fm$ for $q=b$. Thus, in this naive counting,
lattice sizes $(L/a) \gtrsim 90$ and $(L/a) \gtrsim 300$, respectively,
would be needed to keep scaling violations
at the $20\%$ ballpark.\footnote{Evidently, smaller lattices can be used
at the price of introducing systematic uncertainties coming from
chiral extrapolations from heavy pion masses.}
Current state-of-the-art simulations are summarised in~\refig{fig:simulation_landscape}.
The size of the region covered by simulations in the $(a,m_\pi,L)$ parameter space
is determined by the largest computational cost attainable with current supercomputing
resources, which in turn depends on algorithm efficiency and Moore's law. One immediate
conclusion is that, while charm physics is affordable (though still at the price of
working with relatively large pion masses), direct simulation of $b$ quarks is still
very difficult --- indeed, within the accessible region
$(am_b)^2 \gtrsim 1$.
As a consequence, the study of $B$-physics on the lattice requires
input from effective theory, that exploits in various ways the large value of
the $b$-quark mass --- or, more precisely, the small value of the ratio $\lQCD/m_b$ ---
to bypass or assist a direct simulation. The resulting different methodologies will be briefly summarised
below.
Another point worth stressing is that only a fraction of the available ensembles
have been used so far for heavy quark physics. This is partly due to the fact that
the coarser ensembles lead to larger cutoff effects, and partly due to the need
of fully assessing light quark physics before proceeding to the heavy sector.

Prospects for direct simulations of $b$ quarks at their physical mass crucially depend
on the scaling behaviour of algorithmic cost with decreasing $a$.
Until recently, typical cost estimates included a power-law behaviour in $a$,
cf. e.g. the $a^{-6}$ scaling law quoted in~\cite{Giusti:2007hk}.
It has now been
recognised, however, that the performance of common algorithms used in lattice
simulations deteriorates very
rapidly for values of the lattice spacing $a \lesssim 0.05~\fm$, leading
to a surge in autocorrelation times --- especially for quantities very sensitive to
long-distance physics --- which is furthermore essentially insensitive to the
values of sea quark masses~\cite{freezing}.
One particular
consequence is the inability of the algorithm to change topological sector,
which has led to the moniker ``topology freezing'' for this behaviour.\footnote{Initial
evidence mostly relied on simulations of the pure gauge theory, and of
$\NF=2$ QCD with Wilson fermions. Similar findings have recently been reported in
$\NF=2+1$ and $\NF=2+1+1$ simulations with rooted staggered fermions at
lattice spacings below $0.05~\fm$~\cite{Gottlieb:latt15,DeTar:latt15}.}
Another obvious consequence is the impossibility to access directly the $b$-quark mass
region within standard simulation setups.
The existing proposals to avoid the algorithmic critical slowing down for $a \lesssim 0.05~\fm$
involve abandoning the (anti)periodic boundary conditions used in most QCD simulations.
For instance, the introduction of open boundary conditions in Euclidean time --- i.e.
the substitution of the periodic torus for an open-ended cylinder --- has been shown to improve
the scaling of the algorithm considerably~\cite{obc},
and has been incorporated into the latest generation
of large-scale simulations by the Coordinated Lattice Simulations effort~\cite{Bruno:2014jqa}.
The rationale for this approach is that open boundary conditions allow topological
structures to flow in and out of the lattice.
Another very recent proposal~\cite{Mages:2015scv} replaces the periodic torus by a non-orientable
manifold, leading to so-called ``P-periodic'' boundary conditions in Euclidean time.
This results in a similar scaling law for autocorrelation times as open boundary conditions,
albeit autocorrelation times remain significantly larger than in the latter approach.
On the other hand, open boundary conditions break translation invariance in the time
direction and give rise to significant boundary effects, both of which are argued to be absent
with P-periodic boundary conditions.
Finally, in~\cite{Endres:2015yca} a multiscale algorithm is applied to the pure
Yang-Mills theory in a periodic lattice, again showing significant promise in
the reduction of autocorrelations.
While it is still unclear which is the true potential of these new methodologies
in terms of reaching the $a \sim 0.01~\fm$ region, their availability
is a crucial step towards a significant improvement of our control
on the systematics of lattice $B$-physics computations.

\subsection{Approaches to heavy quark physics}

\noindent
As discussed above, the unavailability of LQCD simulations at lattice spacings below
$a\approx 0.05~\fm$ poses a huge challenge for $B$-physics computations, since using
a similar setup as for light quarks will result in extremely large cutoff effects.
Existing approaches to $B$-physics thus rely on input from effective descriptions
of the heavy quark dynamics.
In broad terms, this implies that an expansion in powers of $\lQCD/m_{\rm h}$
(where $m_{\rm h}$ is the mass of the heavy quark) underlies
the procedure, and that some assumptions are made at the field-theoretical level,
including the size of corrections neglected by the truncation of the expansion.
There are two main procedures to perform the expansion:
Heavy Quark Effective Theory (HQET) \cite{hqet},
and Non-Relativistic QCD (NRQCD)~\cite{nrqcd}.
%
HQET provides the correct asymptotic description of QCD
correlation functions in the static limit $|\vp_{\rm h}|/m_{\rm h}\to 0$.
Subleading effects are described by higher dimensional operators, whose coupling constants
are formally of $O((1/m_{\rm h})^n$). The HQET expansion works well for heavy-light systems in which the
heavy-quark momentum is small compared to the mass.
In the static limit the $b$ quark is described by a theory
of a static fermion field coupled to the gauge field, and $b$ propagators
are replaced by Wilson lines in QCD correlation functions; the resulting
theory is renormalisable, and computations can be carried out efficiently. 
NRQCD, on the other hand, is constructed by matching the effective theory to full QCD
order by order in the heavy-quark velocity $v_b^2$ (for heavy-heavy systems)
or in $\lQCD/m_{\rm h}$ (for heavy-light systems), and in powers of $\alpha_{\rm s}$.
Relativistic corrections appear as higher-dimensional operators in the Hamiltonian.
As an effective field theory, NRQCD is only useful with an ultraviolet cutoff of order
$m_{\rm h}$ or less.

\begin{figure}[t!]
\begin{center}
\begin{minipage}[t]{0.24\textwidth}
\vspace{0pt}
\includegraphics[width=\textwidth]{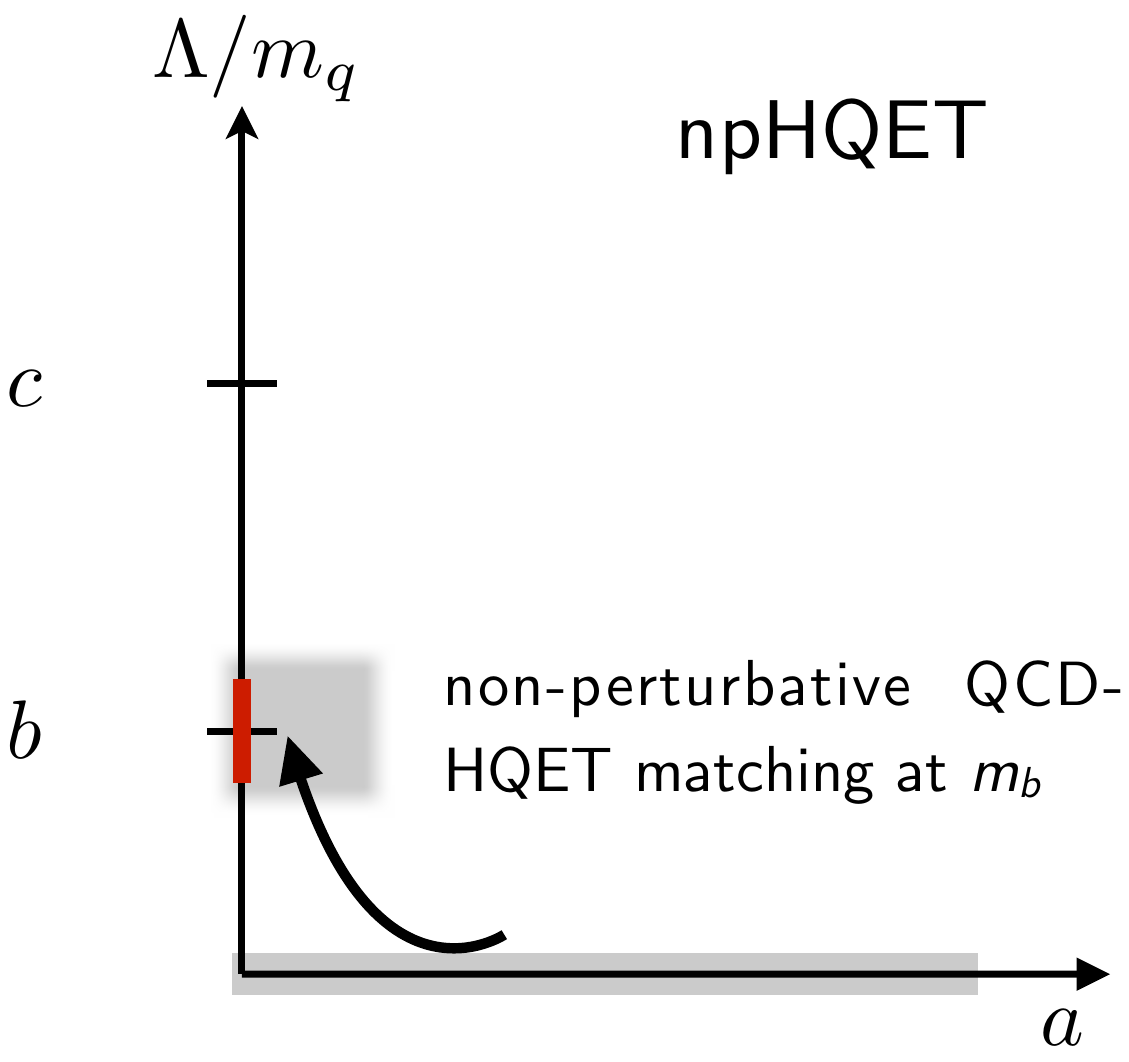}
\end{minipage}
\hspace{0mm}
\begin{minipage}[t]{0.24\textwidth}
\vspace{0pt}
\includegraphics[width=\textwidth]{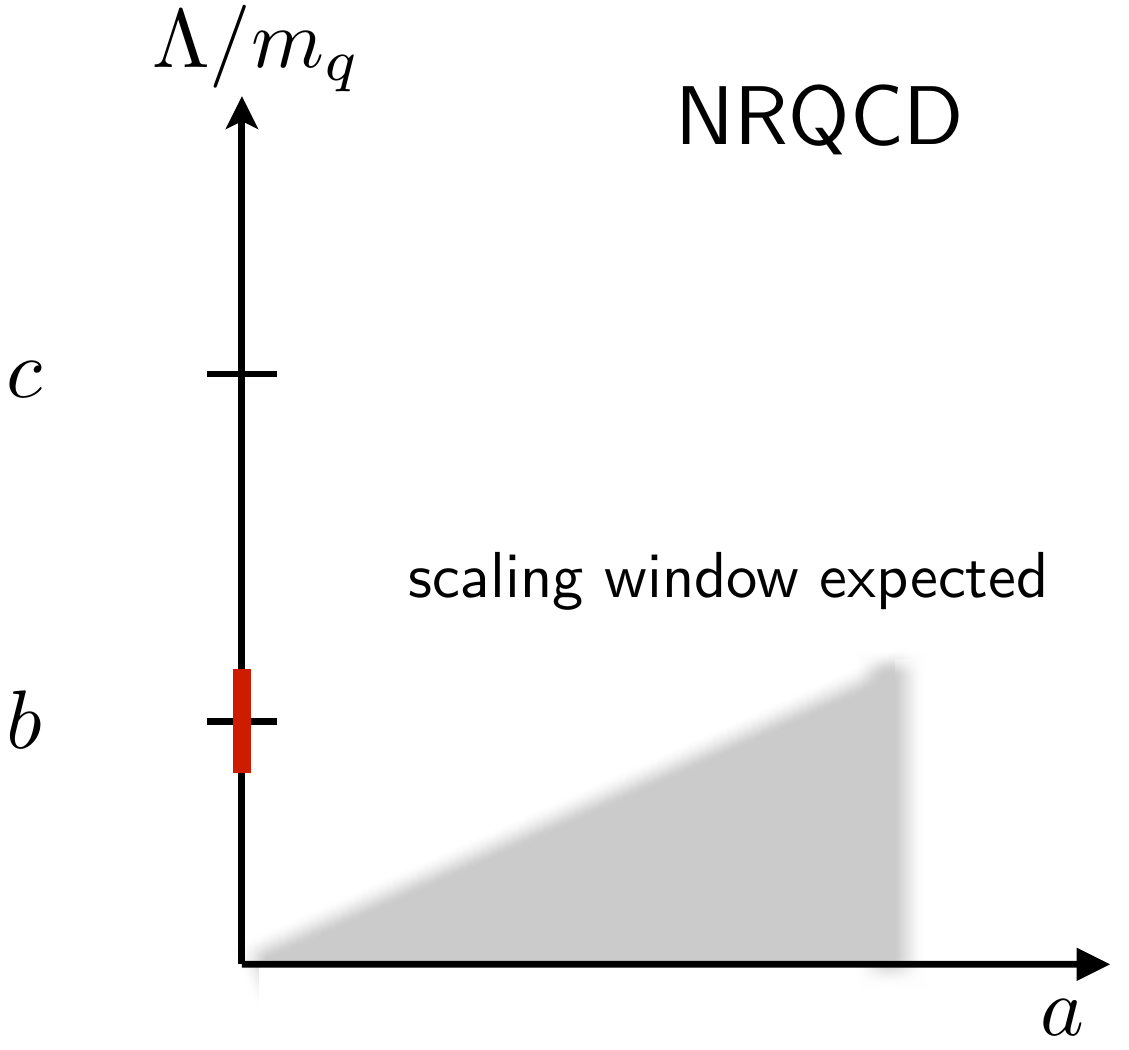}
\end{minipage}
\hspace{0mm}
\begin{minipage}[t]{0.24\textwidth}
\vspace{0pt}
\includegraphics[width=\textwidth]{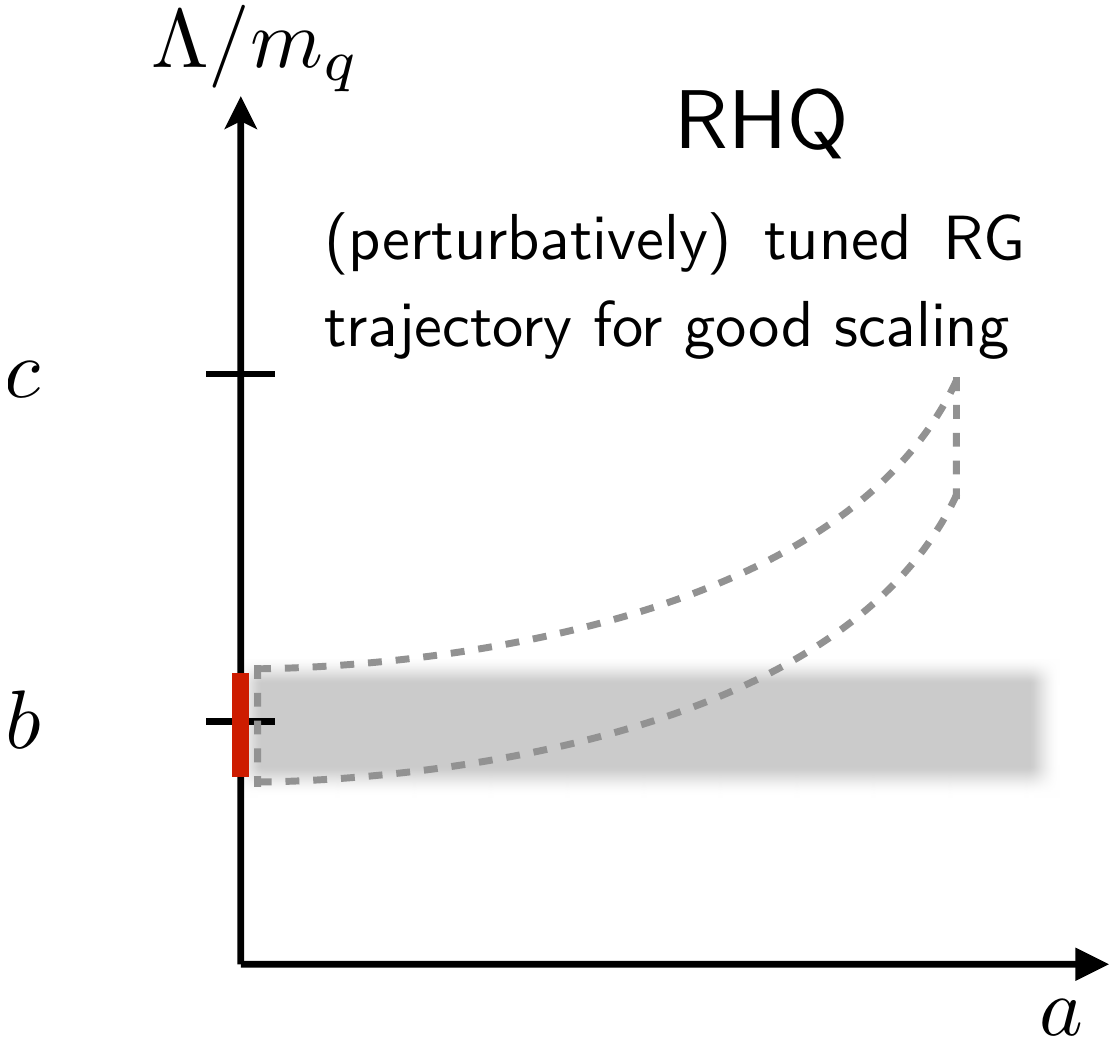}
\end{minipage}
\hspace{0mm}
\begin{minipage}[t]{0.24\textwidth}
\vspace{0pt}
\includegraphics[width=\textwidth]{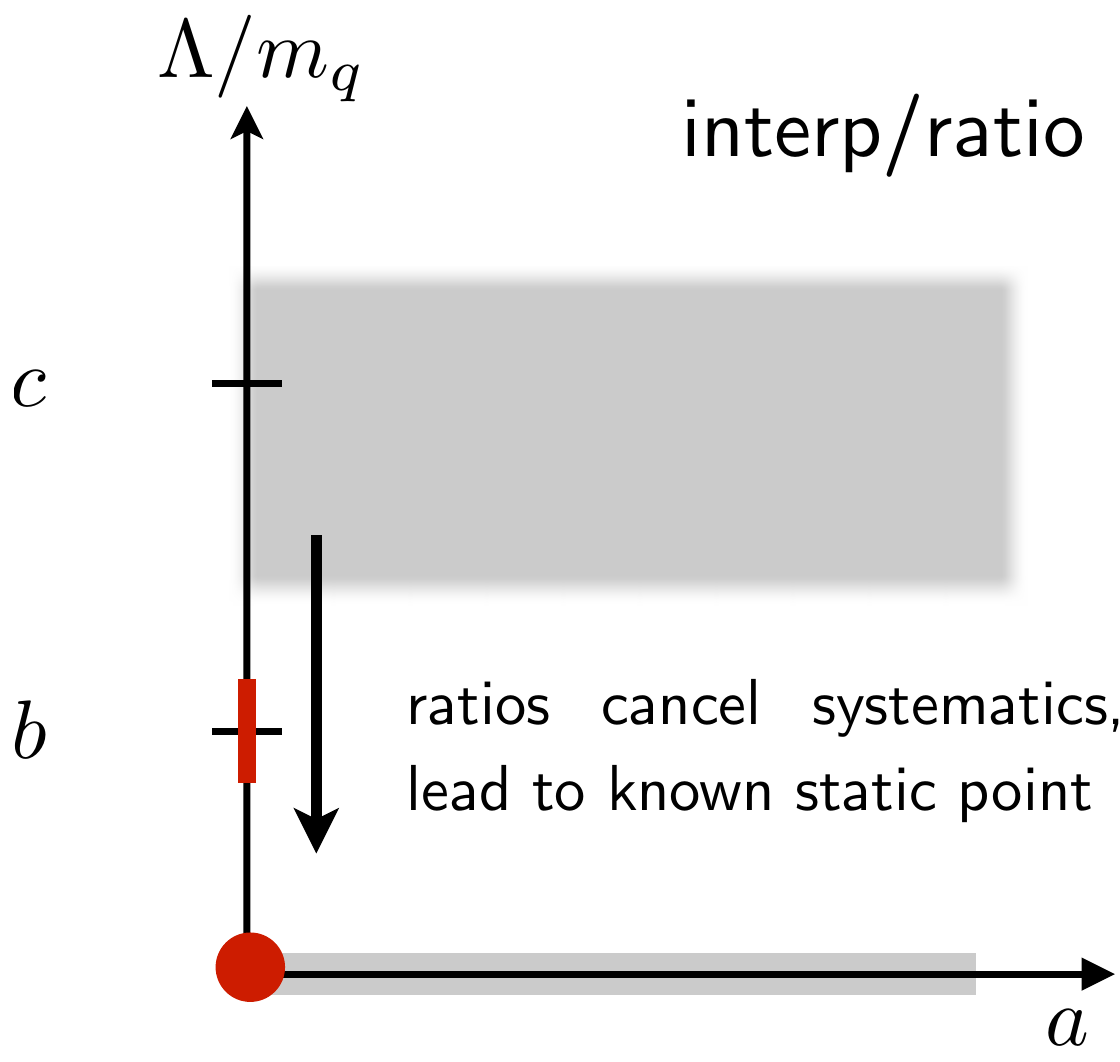}
\end{minipage}
\hspace{0mm}
\end{center}
\vspace{-7mm}
\caption{Cartoon depicting various approaches to heavy quark physics on the lattice
in the plane spanned by the lattice spacing $a$ and the inverse heavy quark mass.}
\label{fig:cartoon}
\end{figure}

The use of effective theory implies that all approaches suffer from systematic uncertainties, although
the extent to which they can be assessed differs widely. In any case, it is crucial
to cross-check the results from different procedures in order to gain confidence
about systematic error estimates.
Below we summarise the main features of each family of currently-used approaches. A cartoon
for each of them is provided in \refig{fig:cartoon}.\footnote{In general, lattice collaborations
stick to one of the methods, as mentioned in the discussion below. Thus, in the coming sections
the formalism employed for $b$ quarks in any given calculation will often not be indicated explicitly.}
%


\subsubsection{Nonperturbative HQET}

\noindent
A framework to treat HQET nonperturbatively was introduced in~\cite{nphqet1}.
The rationale for this approach is based on the observation~\cite{nphqet2}
that,
while $\alpha_{\rm s}(m_{\rm h})$ decreases logarithmically with $m_{\rm h}$,
corrections in the effective theory are power-like in $\lQCD/m_{\rm h}$; therefore,
it is possible that the leading errors in a calculation will be due to the perturbative matching of the action and the currents at a given order $\lQCD/m_{\rm h}$, rather than to the truncation
of the heavy-quark expansion. To eliminate that systematics, the order of the expansion is fixed,
and the matching to QCD is performed nonperturbatively beyond leading order in $\lQCD/m_{\rm h}$.
Higher-dimensional interaction terms in the effective Lagrangian are treated as spacetime volume
insertions into static correlation functions, thus guaranteeing that, at any order
in the heavy-quark expansion, the effective theory is renormalisable.
The implementation used by the ALPHA Collaboration employs two steps, involving separate
sets of simulations. First, the couplings of the effective theory are determined
by matching QCD and HQET nonperturbatively in small physical volumes, where values of
the lattice spacing that allow for a full relativistic treatment of the $b$ quark
are reachable.
Then HQET is simulated in large volumes to compute hadronic observables.
In order to avoid relevant systematics related to the matching in finite volume, the corresponding
length scale $L$ is chosen such that higher orders in $(m_{\rm h}L)^{-1}$ and
$\lQCD/m_{\rm h}$ are of comparable size.

\subsubsection{NRQCD}

\noindent
In a lattice implementation~\cite{nrqcdlatt},
the NRQCD requirement to work with an ultraviolet cutoff
not above $m_{\rm h}$ translates into $am_{\rm h}\gtrsim 1$, which implies that $a$
has to be kept above a minimum value at fixed heavy quark mass, and a continuum limit
cannot be taken. Lattice NRQCD results thus unavoidably retain a cutoff dependence, though
they are expected to be fairly independent of the cutoff within some scaling window,
where physics can be extracted.
One advantage of NRQCD is its ability
to tackle heavy-light and heavy-heavy systems using the same action.
In order to optimise the approach, practical
implementations (in particular, the one employed by the HPQCD Collaboration)
include counterterms in the action that subtract the largest cutoff
effects perturbatively to some fixed order in a simultaneous expansion in powers of $\alpha_{\rm s}$
and $\lQCD/m_{\rm h}$ --- see e.g.~\cite{Gregory:2010gm} for details.
For almost all of the
explored quantities, this usually leads to error budgets where the largest
contribution comes from the $\cO(\alpha_{\rm s}^2)$ uncertainty in the perturbative
matching of operators to full QCD.

\subsubsection{Relativistic heavy-quark actions}

\noindent
Relativistic heavy-quark (RHQ) actions are designed to remove large
$\cO((am_{\rm h})^n,(a|\vp_{\rm h}|)^n)$ cutoff effects by
adjusting the coefficients of suitable higher-dimensional counterterms
via a Symanzik-like procedure.
At fixed lattice spacing, RHQ formulations are
expected to smoothly interpolate between the light-quark and static
limits, which are recovered when $am_{\rm h}\ll 1$ and $am_{\rm h}\gg 1$,
respectively.
A general framework for the approach was developed in~\cite{ElKhadra:1996mp}.
The three most widely used implementations of this idea are the so-called
Fermilab interpretation, commonly employed by FNAL/MILC~\cite{ElKhadra:1996mp}; the RHQ ``Columbia'' formulation developed by
Li, Lin and Christ, in current use by RBC/UKQCD~\cite{Christ:2006us}; and the Tsukuba heavy-quark action introduced in~\cite{Aoki:2001ra}.
All employ an anisotropic action with a Sheikholeslami-Wohlert
term~\cite{Sheikholeslami:1985ij}, and use HQET to constrain the mass dependence of the action and composite operator
improvement coefficients in order to attain the desired scaling properties.
The main difference between Fermilab and Columbia is that, while in the former
case perturbation theory is used
to determine the coefficients, thus leading to truncation errors of
$\cO(\alpha_{\rm s}a|\vp_{\rm h}|,(a|\vp_{\rm h}|)^2)$ at the action level,
in the Columbia action the coefficients are tuned nonperturbatively by
reproducing a set of spectral observables at finite lattice spacing.
The Tsukuba action, on the other hand,
allows for further anisotropies in both dimension-4 and dimension-5
operators in the action, and uses a nonperturbative determination of
the Sheikholeslami-Wohlert coefficient in the massless limit together
with perturbation theory to track the mass dependence of action coefficients.

\subsubsection{Interpolation procedures}

\noindent
In this approach static limit results are combined with full QCD computations,
performed at the largest directly accessible values of the heavy quark mass
(i.e. at or slightly above the charm scale). This allows for an interpolation
in $\lQCD/m_{\rm h}$, that provides the value of the observable of interest at $m_{\rm h}=m_b$.
Such an interpolation can be guided by using HQET predictions for
the heavy quark mass dependence.
A number of specific variants of this general idea have been
proposed~\cite{interp}.
One implementation widely used in recent computations by the ETM Collaboration
and some related efforts, dubbed ``ratio method''~\cite{Blossier:2009hg},
involves considering ratios of values of the observable
of interest computed at different values of the heavy quark mass, keeping the
ratio between consecutive mass values constant. Ratios are built such that their
static limit is trivial. This has a double advantage: uncertainties from the
static value are eliminated, and a number of systematic uncertainties are either
absent or largely cancelled in the ratios --- in particular the bulk of the
large truncation effects~\cite{hqetmatch} induced by a perturbative matching between QCD and HQET.

\section{Leptonic decays}
\label{sec:leptonic}


\noindent
The SM branching fraction for the charged-current-mediated
decay of a $D_{(s)}$ meson is given by
\begin{gather}
\frac{\cB(D_{(s)}\to\ell\nu_\ell)}{\tau_{D_{(s)}}} = 
\frac{\GF^2|V_{cq}|^2}{8\pi}\,f_{D_{(s)}}^2\,m_\ell^2m_{D_{(s)}}\left(1-\frac{m_\ell^2}{m_{D_{(s)}}^2}\right)^2
\label{eq:Dleptonic}
\end{gather}
with $q=d,s$, and where the long-distance QCD contribution is encoded in the decay constant
$f_{D_{(s)}}=\langle 0|\bar c\gamma^\mu\gamma_5 q|D_{(s)}\rangle/(ip^\mu_{D_{(s)}})$.
An experimental measurement of $\cB$ can then be used to determine the CKM matrix element
$|V_{cq}|$, with theory uncertainties (apart from the largely subdominant higher-order OPE terms)
given by the error on $f_{D_{(s)}}$ and the neglect of electromagnetic corrections,
and experimental uncertainties (other than the one on $\cB$ itself)
dominated by the error on the meson lifetime $\tau_{D_{(s)}}$.
The latter two sources of error are $\sim \cO(1\%)$.

An up-to-date summary of experimental results for leptonic charm decay can be found e.g.
in~\cite{Ma:2015tja,Eidelman:2015jya}. The current precision on $\cB(D_s^+\to\mu^+\nu_\mu)$
and $\cB(D_s^+\to\tau^+\nu_\tau)$ quoted by the PDG is 4.5\% and 4.3\%, respectively, while
a recent preliminary measurement of $\cB(D^+\to\mu^+\nu_\mu)$ by BESIII has
significantly improved the precision of this channel, bringing it at 4.5\%~\cite{Agashe:2014kda}.
Meanwhile, prospects for Belle II point at a precision for $D_s\to\tau\nu_\tau$ in the
interval 2.3\%--3.6\% with the long-term expected $50~{\rm ab}^{-1}$ of data, while
for $D_s\to\mu\nu_\mu$ a precision around 1\% is expected~\cite{BelleIIproj}. Note that at that level
of precision electromagnetic corrections many no longer be negligible in this decay.

\begin{figure}[t!]
\begin{center}
\begin{minipage}[t]{0.45\textwidth}
\vspace{0pt}
\includegraphics[width=\textwidth]{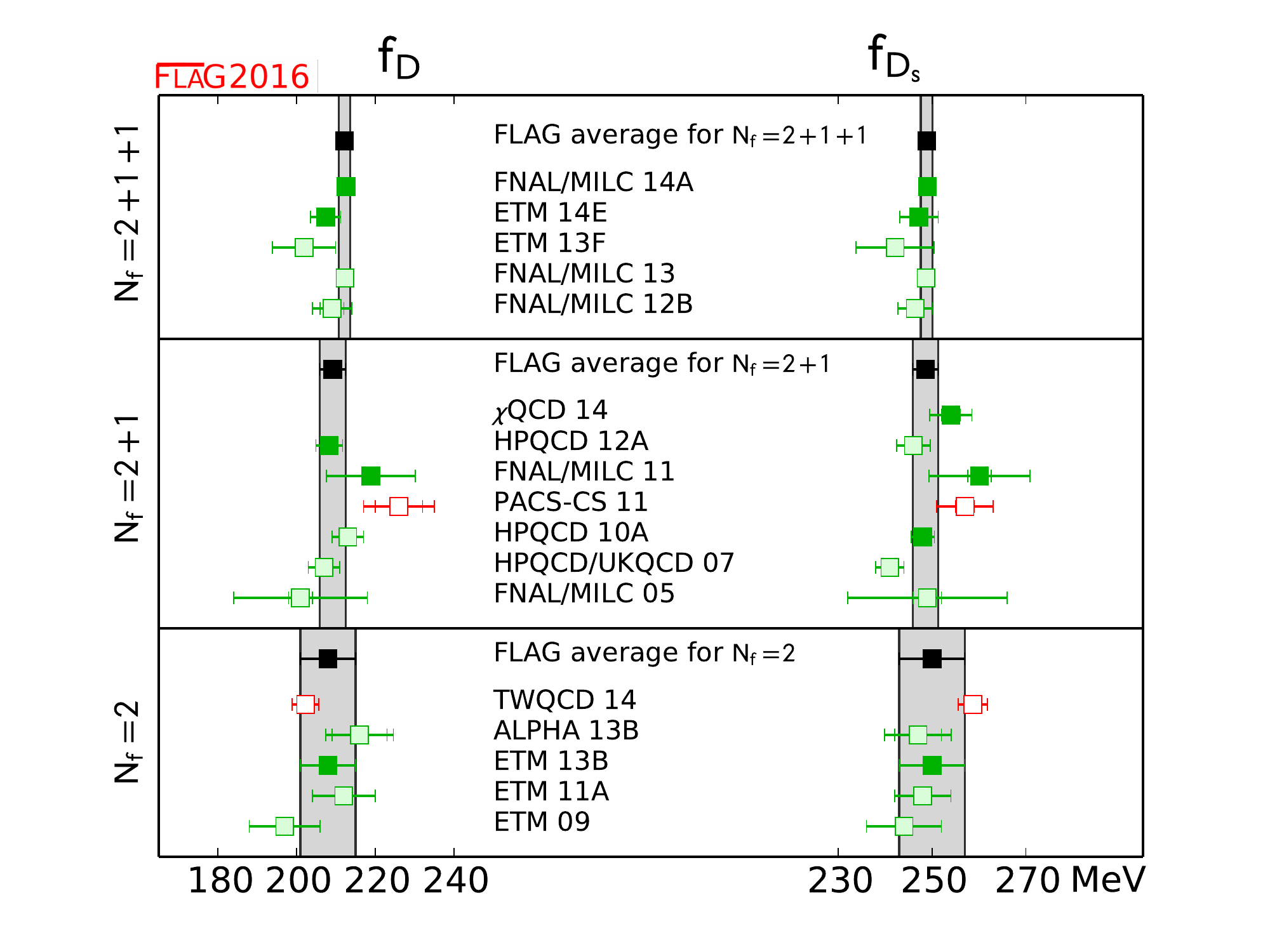}
\end{minipage}
\hspace{10mm}
\begin{minipage}[t]{0.45\textwidth}
\vspace{0pt}
\includegraphics[width=\textwidth]{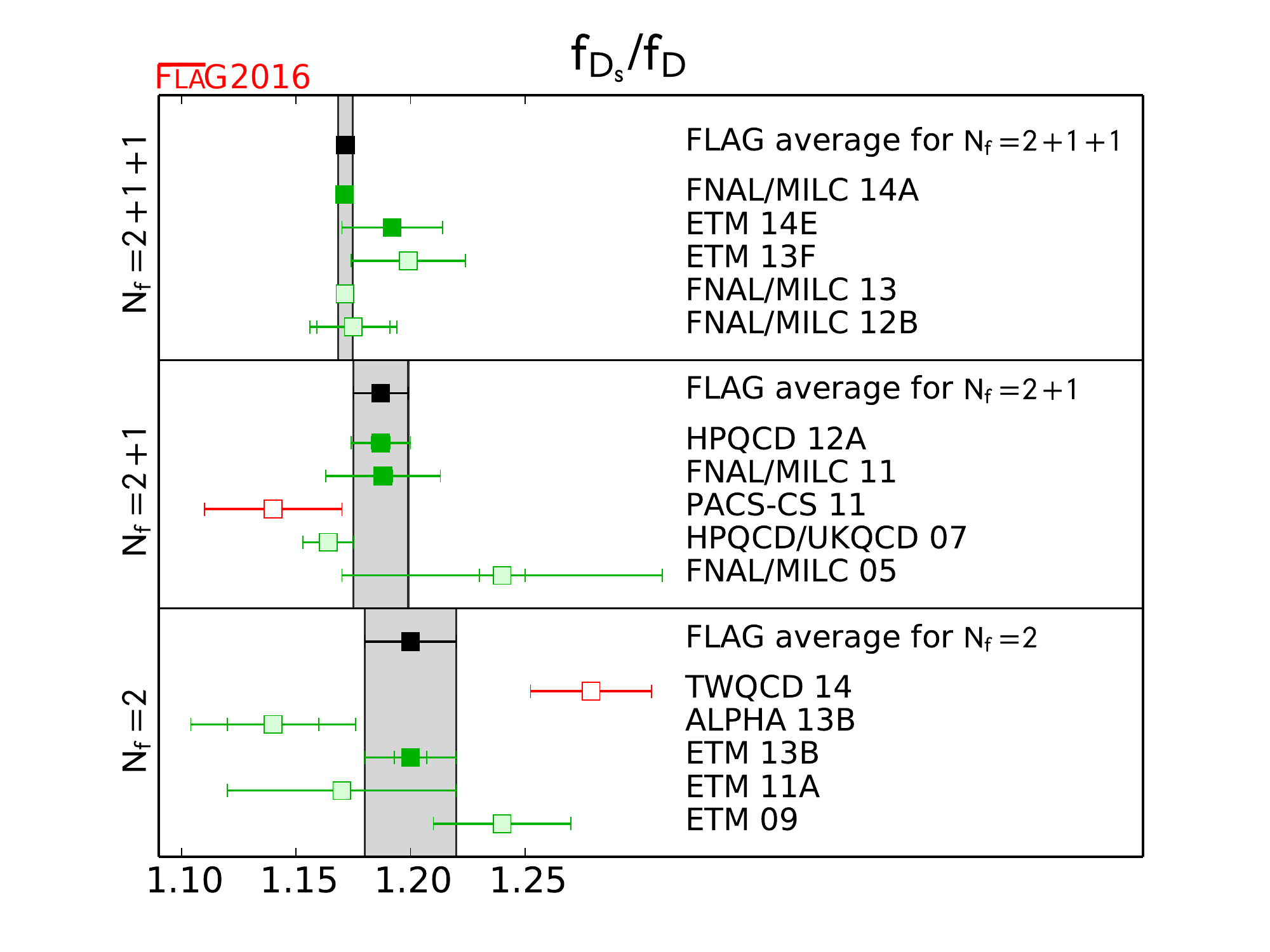}
\end{minipage}
\end{center}
\vspace{-7mm}
\caption{FLAG-3 summary plots for leptonic $D_{(s)}$ decay constants (see~\cite{FLAG3} for a complete list of references).}
\label{fig:Dleptonic}
\end{figure}

The most significant progress in lattice determinations of charm decay constants after FLAG-2
is the extremely precise $\NF=2+1+1$ computation by FNAL/MILC using HISQ quarks~\cite{Bazavov:2014wgs}, which, together
with the final value by ETM~\cite{Carrasco:2013zta}, allows to quote averages with a precision better
than 0.7\% for $f_D$ and 0.5\% for $f_{D_s}$; theoretical improvement on these figures
thus requires a serious attack on electromagnetic corrections. A computation
by $\chi$QCD~\cite{Yang:2014sea} has meanwhile slightly improved the precision of the $\NF=2+1$ determination.
A new result for $\NF=2$ with limited control on systematics has also been provided
by TWQCD~\cite{Chen:2014hva}. The new FLAG averages are given in \ret{tab:fleptonic}
and \refig{fig:Dleptonic}.

The SM branching fraction for the leptonic decay of $B^+$ and $B_c^+$ mesons is given
by \req{eq:Dleptonic}, after replacing meson masses, decay constants, and CKM matrix elements appropriately.
The branching fraction for leptonic $B_s^0$ decay,
which proceeds at one loop in the SM electroweak interaction, is instead
\begin{gather}
\frac{\cB(B_{s}\to\ell\nu_\ell)}{\tau_{B_{s}}} = 
\frac{\GF^2|V_{tb}^*V_{ts}|^2}{\pi}\,f_{B_{s}}^2 \,Y\left(\frac{\alpha}{4\pi\sin^2\theta_W}\right)^2m_{B_s}m_\ell^2
\sqrt{1-4\frac{m_\ell^2}{m_{B_s}^2}}\,,
\end{gather}
where $Y$ is a function that includes NLO QCD and electroweak corrections.
All the decay constants involved are given by the meson-to-vacuum matrix element of an axial current
$\bar b\gamma^\mu\gamma_5 q$; $q=u,s,c$.

\begin{figure}[t!]
\begin{center}
\begin{minipage}[t]{0.45\textwidth}
\vspace{-10pt}
\includegraphics[width=\textwidth]{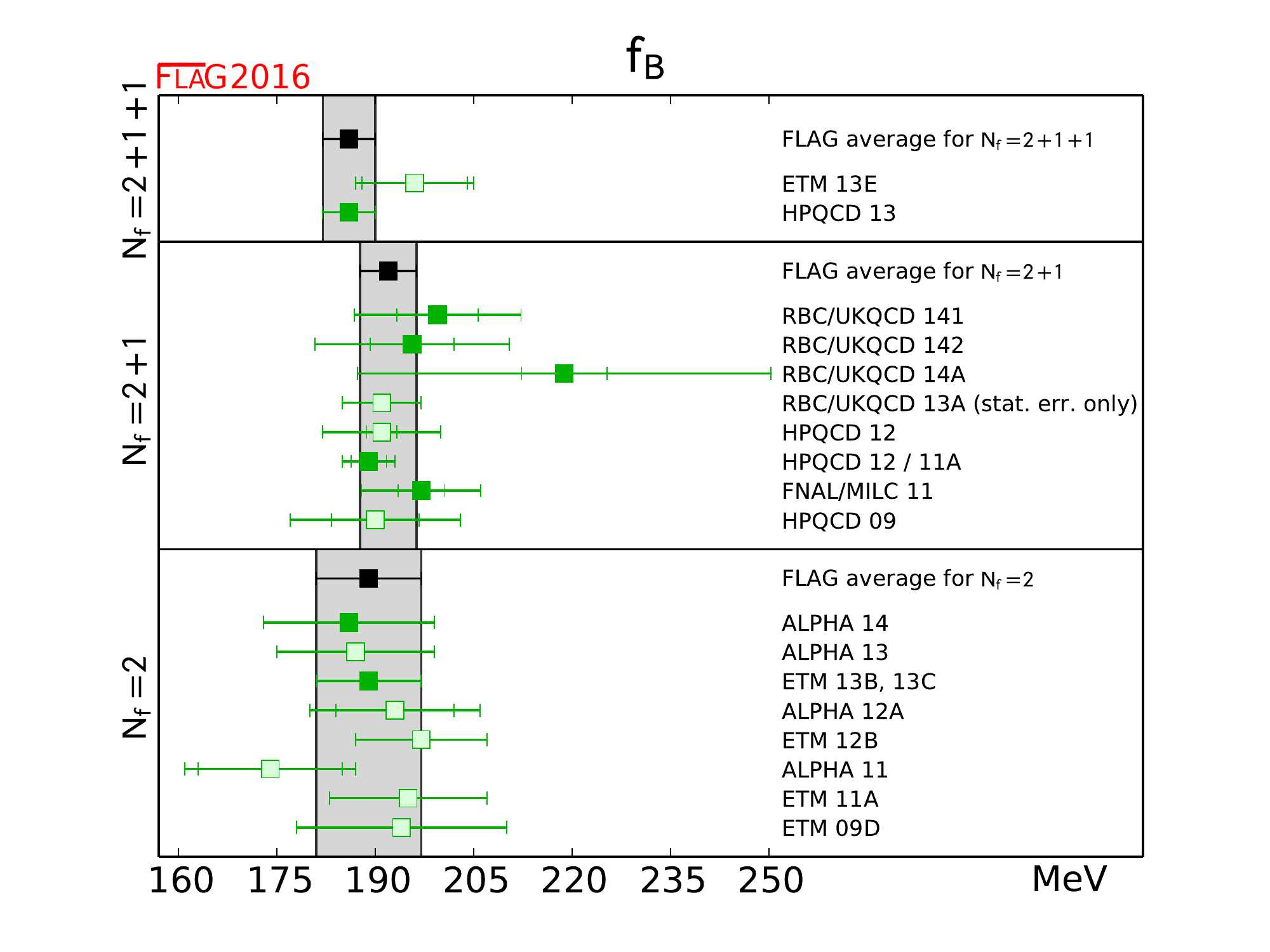}
\end{minipage}
\hspace{10mm}
\begin{minipage}[t]{0.45\textwidth}
\vspace{-10pt}
\includegraphics[width=\textwidth]{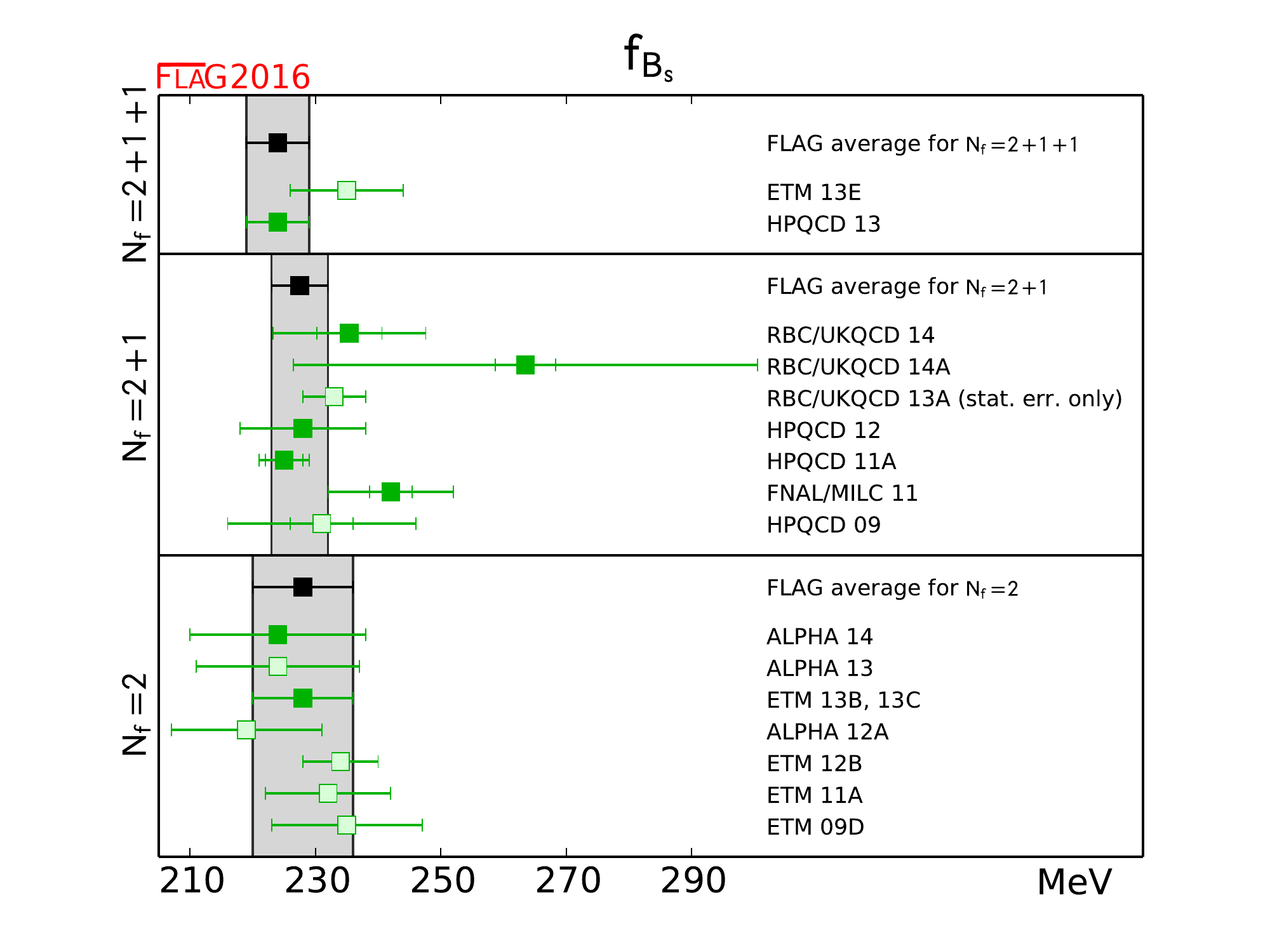}
\end{minipage}
\end{center}
\vspace{-7mm}
\caption{FLAG-3 summary plots for leptonic $B_{(s)}$ decay constants (see~\cite{FLAG3} for a complete list of references).}
\label{fig:Bleptonic}
\end{figure}

\begin{table}[t!]
\begin{center}
\begin{tabular}{l|lll|lll}
\Hline
\multicolumn{1}{c}{$\NF$} & \multicolumn{1}{|c}{$f_D~[\MeV]$} & \multicolumn{1}{c}{$f_{D_s}~[\MeV]$} & \multicolumn{1}{c|}{$f_{D_s}/f_D$} & \multicolumn{1}{c}{$f_B~[\MeV]$} & \multicolumn{1}{c}{$f_{B_s}~[\MeV]$} & \multicolumn{1}{c}{$f_{B_s}/f_B$} \\
\hline
$2+1+1$ & 212.15(1.45) & 248.83(1.27) & 1.1716(32) & 186(4) & 224(5) & 1.205(7) \\
$2+1$   & 209.2(3.3) & 249.8(2.3) & 1.187(12) & 192.0(4.3) & 228.4(3.7) & 1.201(16) \\
$2$     & 208(7) & 250(7) & 1.20(2) & 188(7) & 227(7) & 1.206(23) \\
\Hline
\end{tabular}
\end{center}
\vspace{-5mm}
\caption{FLAG-3 averages for leptonic $D_{(s)}$ and $B_{(s)}$ decay constants.}
\label{tab:fleptonic}
\end{table}

While the tree-level $B^+$ and $B_c^+$ decays
can provide determinations of $|V_{qb}|$ ($q=u,c$), $B_s$ leptonic decay is a powerful probe
of new physics. The current experimental value $\cB(B_s^0\to\mu^+\mu^-)=2.8^{+0.7}_{-0.6}\times 10^{-9}$
\cite{CMS:2014xfa} is well-compatible with the SM prediction.
In the case of tree-level decays, only the $B^+\to\tau^+\nu_\tau$ channel has been measured by both
BaBar~\cite{babarbtau}
and Belle~\cite{bellebtau},
in both cases using different tagging methods.
The uncertainties of these measurements are significantly large, with precisions
of at best 40\%; furthermore, central values tend to be larger than expected from
CKM fits and the more precise semileptonic channels (cf. below).
Uncertainties on lifetime, higher-order OPE and kinematic factors are very small;
the dominant sources of theory uncertainty are the decay constants and
missing electromagnetic corrections, and are however very small compared with
the error on $\cB$ itself. Thus, while this
is an exciting channel in the search for new physics, better experimental precision is
required to establish meaningful comparisons with other exclusive determinations
of $|V_{ub}|$. Belle II projections~\cite{BelleIIproj} expect a dramatic improvement with a 5\% precision
for $\cB(B^+\to\tau^+\nu_\tau)$ with the full $50~{\rm ab}^{-1}$ dataset, while a $5\sigma$
measurement of $\cB(B^+\to\mu^+\nu_\mu)$ is also foreseen.\footnote{The leptonic decay of the
charmed $B_c^+$ meson has not
been measured experimentally yet, but a SM prediction exists based on the HPQCD computation of the
relevant decay constant~\cite{McNeile:2012qf}.}

On the theory side, relatively little progress has been made after the FLAG-2
review concerning determinations of $f_B$ and $f_{B_s}$: new results include
the final npHQET $\NF=2$ ALPHA values~\cite{Bernardoni:2014fva}, new $\NF=2+1$ results from RBC/UKQCD
with relatively large errors~\cite{Christ:2014uea,Aoki:2014nga},
using a RHQ formulation for
the $b$ quark, and preliminary $\NF=2+1+1$ results by ETM~\cite{Carrasco:2013naa} using their ratio method.
The current situation is summarised in \ret{tab:fleptonic}
and \refig{fig:Bleptonic}. While the uncertainties are generally larger
than for charm decay, due to the complications related to the treatment of
$b$ quarks, the precision ballpark is already at the few percent level,
thus ahead of the foreseeable experimental precision.


\section{$B^0$--$\bar B^0$ mixing}
\label{sec:mixing}

\begin{figure}[t!]
\begin{center}
\begin{minipage}[t]{0.45\textwidth}
\vspace{0pt}
\includegraphics[width=\textwidth]{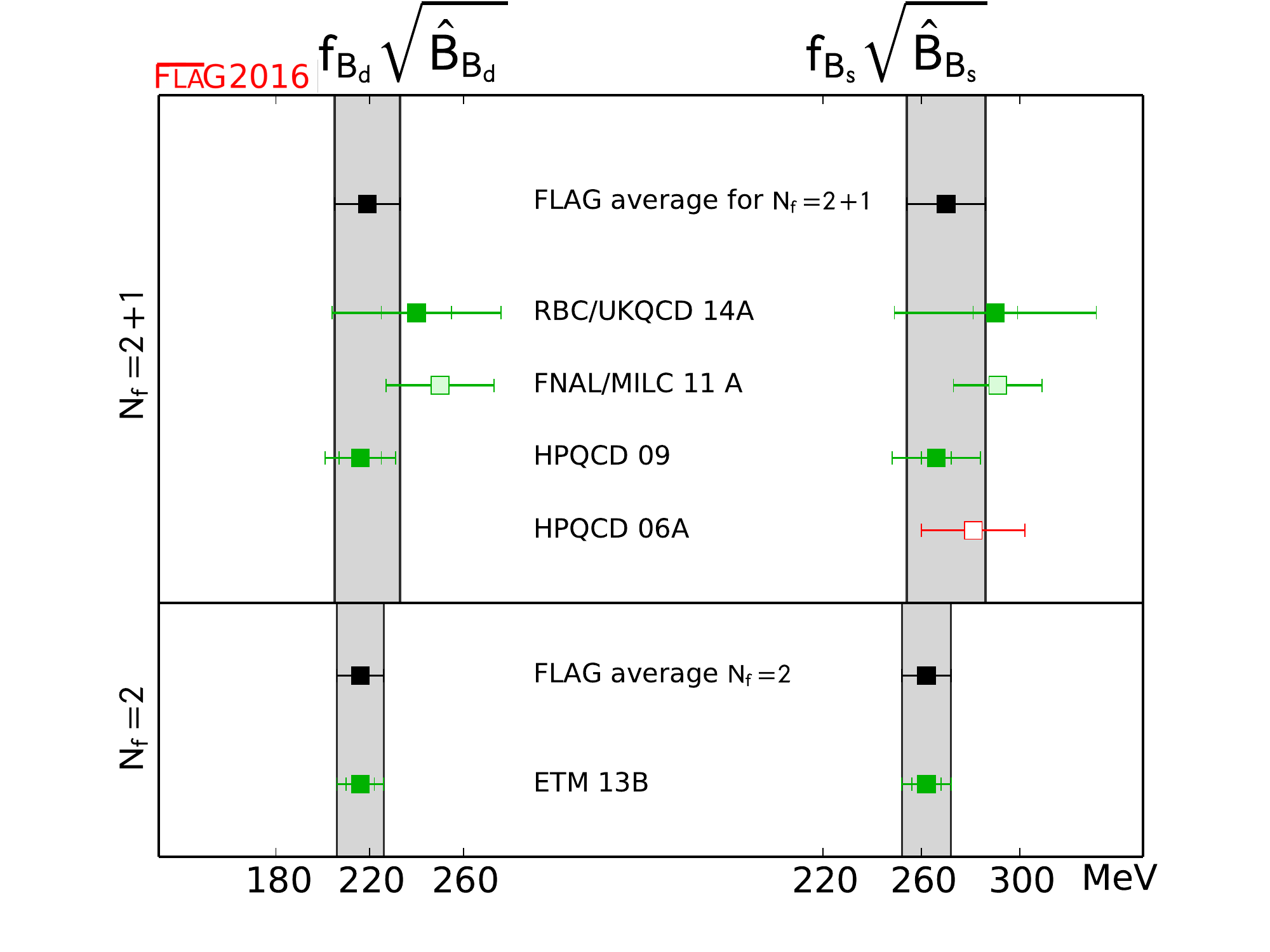}
\end{minipage}
\hspace{10mm}
\begin{minipage}[t]{0.45\textwidth}
\vspace{0pt}
\includegraphics[width=\textwidth]{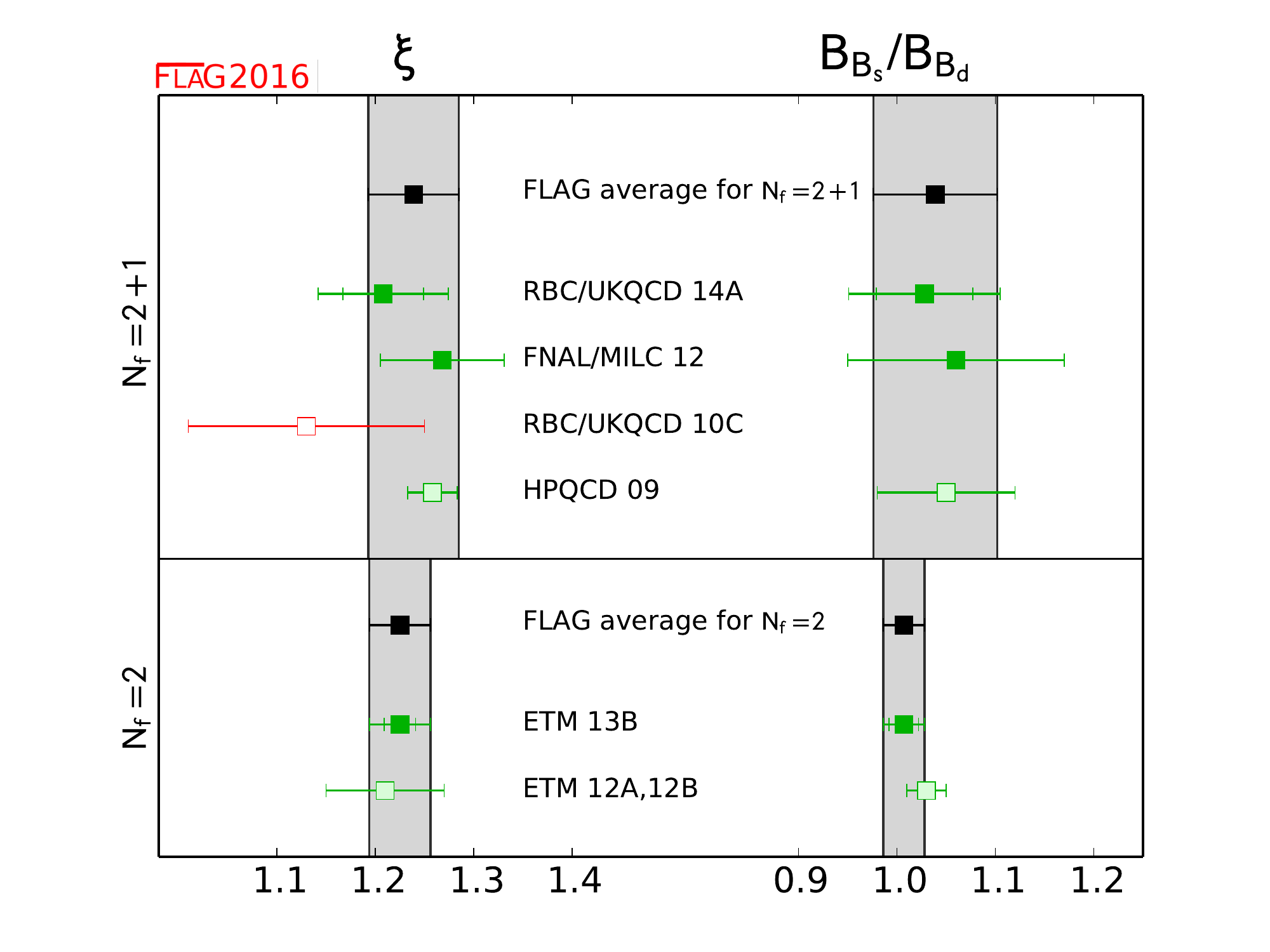}
\end{minipage}
\end{center}
\vspace{-7mm}
\caption{FLAG-3 summary plots for $B$-meson bag parameters (see~\cite{FLAG3} for a complete list of references).}
\label{fig:Bmixing}
\end{figure}

\begin{table}[t!]
\begin{center}
\begin{tabular}{l|ll|ll|ll}
\Hline
\multicolumn{1}{c}{$\NF$} &
\multicolumn{1}{|c}{$f_{B_d}\sqrt{\orgi{B}_{B_d}}~[\MeV]$} & \multicolumn{1}{c}{$f_{B_s}\sqrt{\orgi{B}_{B_s}}~[\MeV]$} &
\multicolumn{1}{|c}{$\orgi{B}_{B_d}$} & \multicolumn{1}{c}{$\orgi{B}_{B_s}$} &
\multicolumn{1}{|c}{$\xi$} & \multicolumn{1}{c}{$\orgi{B}_{B_s}/\orgi{B}_{B_d}$} \\
\hline
$2+1$   & 219(14) & 270(16) & 1.26(9) & 1.32(6) & 1.239(46) & 1.039(63) \\
$2$     & 216(10) & 262(10) & 1.30(6) & 1.32(5) & 1.225(31) & 1.007(21) \\
\Hline
\end{tabular}
\end{center}
\vspace{-5mm}
\caption{FLAG-3 averages for $B$-meson bag parameters.}
\label{tab:Bmixing}
\end{table}

\noindent
Neutral $B$-meson mixing is induced in the SM, to lowest order in the electroweak theory,
through one-loop box diagrams, resulting in an effective $\Delta B=2$ weak Hamiltonian of the form
\begin{gather}
H_{\rm w}^{\Delta B=2} = \frac{\GF^2M_W^2}{16\pi^2}S_0\left(\frac{m_t^2}{M_W^2}\right)\eta_{2B}\left[(V_{td}^*V_{tb})^2 Q_1^d\,+\,(V_{ts}^*V_{tb})^2 Q_1^s\right]\,, \quad
Q_1^q = (\bar b_{\rm L}\gamma_\mu q_{\rm L})(\bar b_{\rm L}\gamma_\mu q_{\rm L})\,,
\end{gather}
where electroweak and high-energy QCD corrections are contained in the Inami-Lim function $S_0$
and the factor $\eta_{2B}$, respectively.
$\orgi{Q}_1^q$ indicates the renormalisation group invariant (RGI) operator, obtained
from the renormalised operator at scale $\mu$ by calculating the anomalous dimension and beta
function appearing in the renormalisation
group equations for the operator and the strong coupling constant,
\begin{gather}
\mu\frac{\partial}{\partial\mu}\obar{Q}(\mu)=\gamma(\gbar(\mu))\obar{Q}(\mu)\,, \qquad
\mu\frac{\partial}{\partial\mu}\gbar(\mu)=\beta(\gbar(\mu))\,,
\end{gather}
with perturbative expansions $\gamma(g) = -\gamma_0g^2+\ldots$ and $\beta(g) = -b_0g^3+\ldots$,
and computing
\begin{gather}
\orgi{Q}=\left[\frac{\gbar^2(\mu)}{4\pi}\right]^{-\frac{\gamma_0}{2b_0}}\exp\left\{-\int_0^{\gbar(\mu)}
\dif g \left[\frac{\gamma(g)}{\beta(g)}-\frac{\gamma_0}{b_0g}\right]\right\}\,\obar{Q}(\mu)\,.
\end{gather}
Long-distance QCD contributions are encoded in the bag parameters
\begin{gather}
\orgi{B}_{B_q}=\frac{\langle\bar B_q^0|\orgi{Q}_1^q|B_q^0\rangle}{\frac{8}{3}f_{B_q}^2m_{B_q}^2}\,, \qquad\qquad
\xi^2 = \frac{f_{B_s}^2 \orgi{B}_{B_s}}{f_{B_d}^2 \orgi{B}_{B_d}}\,,
\label{eq:Bpar}
\end{gather}
where we have also defined the ratio $\xi$, that will be discussed below.
Note that $\orgi{B}_{B_q}$ and $\xi$ are scale- and renormalisation-scheme-independent by construction.

A non-zero mixing amplitude results in mass differences between the CP eigenstates of the neutral
meson system, for which the SM prediction is
\begin{gather}
\Delta m_q = \frac{\GF^2M_W^2m_{B_q}}{6\pi^2}|V_{tq}^*V_{tb}|^2S_0\left(\frac{m_t^2}{M_W^2}\right)\eta_{2B}\,f_{B_q}^2\orgi{B}_{B_q}\,.
\end{gather}
These quantities are experimentally measurable to high precision --- current PDG
averages~\cite{Agashe:2014kda} quote 0.6\% and 0.1\% for $\Delta m_d$
and $\Delta m_s$, respectively.
Another interesting observable is the ratio
$\Delta m_s/\Delta m_d$, where short-distance effects cancel and the long-distance QCD
contribution is encoded in the ratio $\xi$ in~\req{eq:Bpar},
which can be computed on the lattice to significantly better precision than the individual
$f_{B_q}^2\orgi{B}_{B_q}$.
Using a measurement of $\Delta m_q$ and a computation of $f_{B_q}^2\orgi{B}_{B_q}$ it is then
possible to determine $|V_{tq}^*V_{tb}|^2$, and feed it to a unitarity triangle
analysis on the $\bar\rho-\bar\eta$ plane of Wolfenstein parameters, where, given the values
of $|V_{tq}|$, it constrains the position of the triangle apex
to lie on a circumference.
Since the current precision on the knowledge of the relevant combination
of CKM moduli is in the 7--8\% ballpark, this sets the precision target on $\orgi{B}_{B_q}$
to avoid dominant theory uncertainties.

Unfortunately, relatively few results exist yet for $B$-meson bag parameters and/or their ratios; and the
only update from the FLAG-2 review is an $\NF=2+1$ RBC/UKQCD computation using a RHQ
treatment for the $b$ quark~\cite{Aoki:2014nga}, with however significantly larger errors than the pre-existing
HPQCD~\cite{Gamiz:2009ku} and FNAL/MILC~\cite{Bouchard:2011xj} results. Together with the now-published ETM values
for $\NF=2$~\cite{Carrasco:2013zta}, this results in the landscape illustrated by \refig{fig:Bmixing}
and \ret{tab:Bmixing}.

\section{Semileptonic decays}
\label{sec:SL}

\subsection{$D$-meson decays}

\noindent
The SM differential rate for $D\to P\ell\nu_\ell$ decay with $P=\pi,K$
($q=d,s$) is given by
\begin{gather}
\begin{split}
\frac{\dif\Gamma(D\to P\ell\nu_\ell)}{\dif q^2} = \frac{\GF^2|V_{cq}|^2}{24\pi^3}\,
\frac{(q^2-m_\ell^2)^2\sqrt{E_P^2-m_P^2}}{q^4m_D^2}\Bigg[
&\left(1+\frac{m_\ell^2}{2q^2}\right)m_D^2(E_P^2-m_P^2)|f_+(q^2)|^2 \\[-1.5ex]
&~+\frac{3m_\ell^2}{8q^2}(m_D^2-m_P^2)^2|f_0(q^2)|^2
\Bigg]\,,
\end{split}
\label{eq:Dslrate}
\end{gather}
where $E_P$ is the energy of the outgoing meson, $q$ is the total four-momentum transferred to the lepton pair,
and the vector and scalar form factors $f_{+,0}$, normalised such that $f_+(0)=f_0(0)$,
parametrise the hadronic matrix element of the relevant charged current,
\begin{gather}
\langle P|\bar q\gamma^\mu c|D\rangle = f_+(q^2)\left[p_D^\mu+p_P^\mu-\frac{m_D^2-m_P^2}{q^2}\,q^\mu\right]
\,+\,f_0(q^2)\,\frac{m_D^2-m_P^2}{q^2}\,q^\mu\,.
\label{eq:ffs}
\end{gather}
The contribution to \req{eq:Dslrate} coming from $f_0$ is in practice negligible for $\ell=e,\mu$.
Since, on the other hand, there are no experimental results for $\tau$ channels either in
$D^+$ or $D^0$ decay,\footnote{It is worth mentioning that, in the case of the $D_s$ meson,
the only measured semileptonic mode with a pseudoscalar meson in the final state
is $D_s\to K^0e^+\nu_e$, and the precision of the measurement
is much poorer than in the case of the $D^{+,0}$ semileptonic modes. These decays have therefore
received little attention.}
lattice collaborations have focused on computations of the vector form factor $f_+$,
which in turn allows to extract the value of the CKM matrix elements $|V_{cd}|$ and $|V_{cs}|$.
While the kinematically allowed interval of values for $q^2$ is quite broad, all published
lattice computations only provide the form factor at zero momentum transfer $f_+(0)$,
which is sufficient to extract the CKM by comparing the form-factor normalisation or
matching the total branching fraction for the process.
Until recently, the PDG averages for the total branching fractions of these processes were
dominated by the BaBar~\cite{babardsl}
and CLEO-c~\cite{Besson:2009uv}
measurements. New results
from BESIII~\cite{Ma:2015tja} will however result in a significant improvement. In particular, they
provide very accurate determinations of the form-factor shapes (cf. Fig.~4 in \cite{Ma:2015tja}), that
substantially improve on the CLEO-c, BaBar, and Belle existing results for the latter.

There is a marked paucity of lattice results for the form factors: both in FLAG-2 and FLAG-3
the only publications contributing to averages are the HPQCD $\NF=2+1$ works~\cite{hpqcddsl}.
New work is however being intensely pursued by other collaborations, and progress has been presented
at this conference: they include the high-precision $\NF=2+1+1$ computations by ETM~\cite{Lami:latt15,Carrasco:2016kpy}
and FNAL/MILC~\cite{Primer:latt15}, and preliminary $\NF=2+1$ results from JLQCD~\cite{Suzuki:latt15}. As we will
discuss later, these are very interesting channels, both on their own phenomenological right, and
for the purpose of understanding the systematic uncertainties involved in the description
of the $q^2$ dependence of form factors in semileptonic heavy meson decay.

\subsection{$B_{(s)} \to D^{(*)}_{(s)}\ell\nu$ decays}

\begin{figure}[t!]
\begin{center}
\setbox1=\hbox{\includegraphics[width=0.45\textwidth]{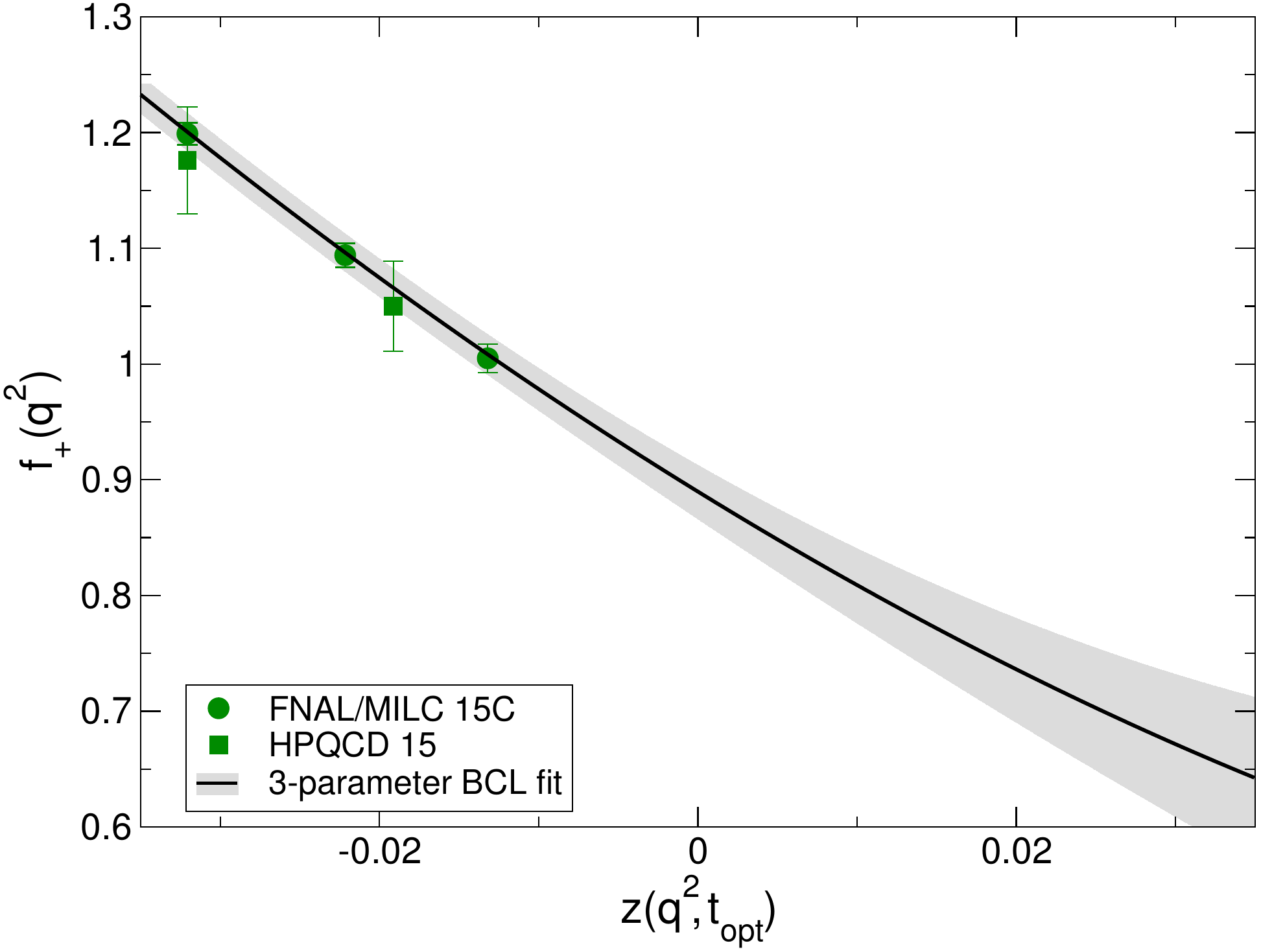}}
\setbox2=\hbox{\includegraphics[width=0.45\textwidth]{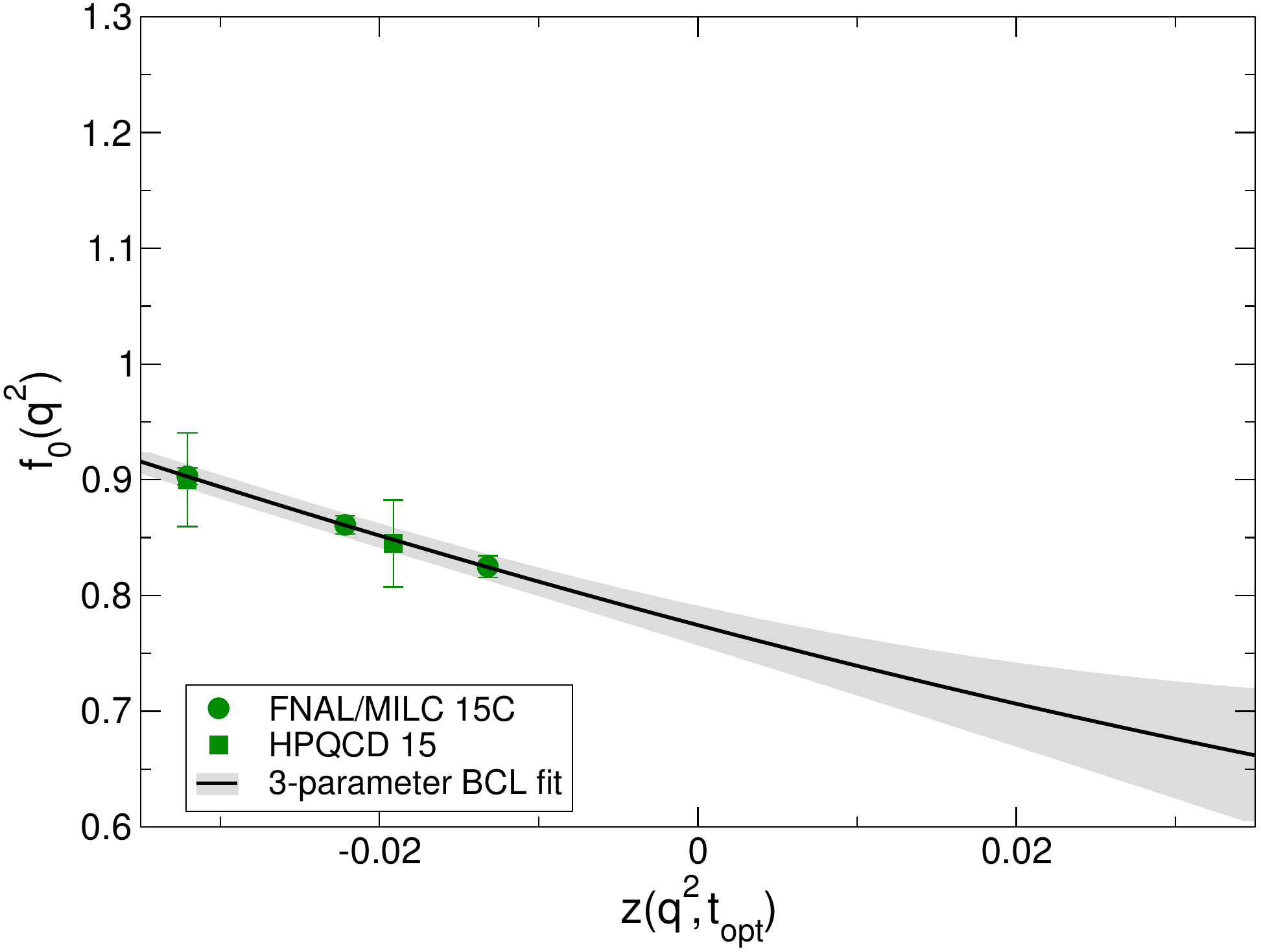}}
\begin{minipage}{0.45\textwidth}
\includegraphics[width=\textwidth]{./figs/fp_BD_latt-eps-converted-to.pdf}
\end{minipage}
\begin{minipage}{0.45\textwidth}
\includegraphics[width=\textwidth]{./figs/f0_BD_latt-eps-converted-to.pdf}
\end{minipage}
\begin{minipage}{0.45\textwidth}
   \llap{\makebox[\wd1][l]{\raisebox{55mm}{
   \hspace{-66mm}\includegraphics[width=0.15\textwidth]{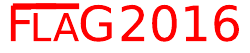}
   }}}
\end{minipage}
\begin{minipage}{0.45\textwidth}
   \llap{\makebox[\wd2][l]{\raisebox{55mm}{
   \hspace{4mm}\includegraphics[width=0.15\textwidth]{./figs/FLAG_Logo-eps-converted-to.pdf}
   }}}
\end{minipage}
\vspace{-3mm}
\caption{FLAG-3 BCL fits (grey band) for form factors $f_+(q^2)$
(left) and $f_0(q^2)$ (right) for $B \to D\ell\nu$, plotted versus $z(q^2)$.
The $w=1$ point corresponds to $z\simeq -0.0323$.
(See~\cite{FLAG3} for a complete list of references and a discussion of dataset
and fit details.)}
\label{fig:BtoD}
\end{center}
\end{figure}

\noindent
The most relevant exclusive modes used to determine $|V_{cb}|$ are the CKM-favoured
$B_{(s)}$ semileptonic decays with a $D_{(s)}$ or a $D^*_{(s)}$ meson in the final state. Their SM rates
for $\ell=e,\mu$ are given by\footnote{In the following we drop $_{(s)}$ subscripts; the channel being discussed will be clear by context, or explicitly indicated.}
\begin{align}
    \frac{\dif\Gamma(B\to D \ell\nu_\ell)}{\dif w} & = 
        \frac{\GF^2 m^3_{D}}{48\pi^3}(m_B+m_{D})^2(w^2-1)^{3/2}  |\eta_\mathrm{EW}|^2|V_{cb}|^2 |\cG(w)|^2,
    \label{eq:BtoD} \\
    \frac{\dif\Gamma(B\to D^* \ell\nu_\ell)}{\dif w} & = 
        \frac{\GF^2 m^3_{D^*}}{4\pi^3}(m_B-m_{D^*})^2(w^2-1)^{1/2}  |\eta_\mathrm{EW}|^2|V_{cb}|^2\chi(w)|\cF(w)|^2 ,
    \label{eq:BtoDstar}
\end{align}
where $w=p_B\cdot p_{D^{(*)}}/|p_B\cdot p_{D^{(*)}}|$ is the recoil parameter, $\eta_\mathrm{EW}$
contains electroweak loop corrections, and $\chi$ is a function of $w$ and meson masses
that fulfills $\chi(1)=1$. In the case of the rate for $B\to D\ell\nu$, \req{eq:Dslrate}
can be alternatively used with appropriate replacements in mass and CKM factors;
the form factor $\cG$ appearing in \req{eq:BtoD}
is related to the standard vector form factor (cf. \req{eq:ffs}) by $f_+(q^2)=\quart(1+\frac{m_B}{m_D})\cG(w)$,
with $q^2=m_B^2+m_D^2-2wm_Bm_D$. Also of interest are the ratios of the total branching fractions
in the $\tau$ channel relative to those in the light lepton channels, where
CKM factors and electroweak corrections cancel, and which are strongly sensitive
to the scalar form factors,
\begin{gather}
R(D^{(*)}) = \frac{\cB(B\to D^{(*)} \tau\nu_\tau)}{\cB(B\to D^{(*)} \ell\nu_\ell)}\,.
\end{gather}
Evidently, from the theory point of view the $D^*$ channel poses greater technical
complications than the $D$: the $D^*$ is unstable, which in principle requires a non-trivial procedure
to extract the physical amplitude from an Euclidean correlation function;\footnote{Note that this is
very sensitive to light quark masses, since the $D^*$ mass is very close to the $D\pi$ threshold. In physical kinematics the width
is in any case small enough so that its effect can likely be neglected at the current level of precision.}
and a more complicated kinematics is involved.

Experimental measurements are, on the other hand, significantly more precise in the $D^*$ channel.\footnote{The precision on the world average for the $B^0\to D^-\ell\nu$ total branching fraction
quoted by the latest HFAG report is 4.5\%, while for $B^0\to D^{*-}\ell\nu$ is 2.2\% --- see~\cite{Amhis:2014hma}
for a full discussion.}
The precision in $B\to D\ell\nu$ has however been significantly improved by a recent update
by Belle~\cite{Glattauer:2015teq}, which is now the most precise measurement available.
In the case of $\tau$ channels and the resulting values for $R(D^{(*)})$,
the experimental precision is again significantly better in
the $D^*$ channel (9.0\% vs. 16.5\% quoted in~\cite{Lees:2013uzd}), which has been further reinforced by
a recent update by Belle~\cite{Huschle:2015rga} and a new measurement by LHCb~\cite{Aaij:2015yra}. The Belle II
projection is to improve the precision on both observables by a factor of 4~to~5 with
the full $50~{\rm ab}^{-1}$ dataset~\cite{BelleIIproj}. In the case of the channels with light
leptons, the error is expected to be halved.

Lattice results with dynamical fermions had focused until recently on the form factors
at zero recoil, and FLAG-2 reported $\NF=2+1$ averages for $\cG^{B\to D}(1)$ and $\cF^{B\to D^*}(1)$,
based on FNAL/MILC results~\cite{fnaloldbd};
$\NF=2$ results for $\cG^{B\to D}(1)$ and $\cG^{B_s\to D_s}(1)$ obtained in~\cite{Atoui:2013zza} have also
been published since then, and enter averages in FLAG-3.
Recently, however, HPQCD~\cite{Na:2015kha} and FNAL/MILC~\cite{Lattice:2015rga} published first detailed $\NF=2+1$ studies
of form-factor shapes for both $f_+$ and $f_0$, which in the latter case supersedes previous
determinations of $\cG^{B\to D}(1)$. Meanwhile, FNAL/MILC has also updated their $\cF^{B\to D^*}(1)$
value~\cite{Bailey:2014tva}. This has led to a marked increase in the control over systematic uncertainties, resulting
in more accurate exclusive determinations of $|V_{cb}|$ (see~\res{sec:CKM}). The form factor determinations
by HPQCD and FNAL/MILC can be averaged into a single function of $q^2$ (or $w$), using
e.g. the FLAG-recommended BCL fit ansatz for the momentum transfer dependence (see~\res{sec:percent});
the result is illustrated in \refig{fig:BtoD}. Remarkably, as noted in~\cite{Lattice:2015rga}, the $B\to D$
vector form-factor shape obtained for $\NF=2+1$ is well-compatible with the quenched
result from~\cite{deDivitiis:2007otp}.

The availability of new results for the scalar form factor have also allowed for much
more precise SM predictions for $R(D)$, with respect to the pre-existing FNAL/MILC value~\cite{Bailey:2012jg};
the resulting FLAG-3 average is
\begin{gather}
R(D)=0.300(8)\,.
\end{gather}
It is worth stressing that no lattice-based computation of $R(D^*)$ is currently
available. The commonly quoted SM prediction for $R(D^*)$~\cite{Fajfer:2012vx} is based on a
phenomenological analysis that takes as input the experimental values for the form
factors at zero recoil and the $w$ dependence in the heavy-quark limit,
and estimates hadronic uncertainties using higher-order perturbative and power corrections
to the latter. The resulting uncertainties largely cancel in $R(D^*)$, leading to an
error much smaller than the one quoted for lattice determinations of $R(D)$.

\subsection{$B \to \pi\ell\nu$ and $B_s\to K\ell\nu$ decays}

\noindent
The CKM-suppressed decay $B\to\pi\ell\nu_\ell$ is the most relevant exclusive channel
for the determination of $|V_{ub}|$. Both it and the very similar $B_s\to K\ell\nu$
decay have SM differential rates given by \req{eq:Dslrate}, replacing $D\to B_{(s)}$ and $V_{cq}\to V_{ub}$;
the relevant charged flavour current is $\bar u\gamma_\mu b$, which can again be parametrised
by vector and scalar form factors as in~\req{eq:ffs}.

\begin{figure}[t!]
\begin{center}
\setbox1=\hbox{\includegraphics[width=0.45\textwidth]{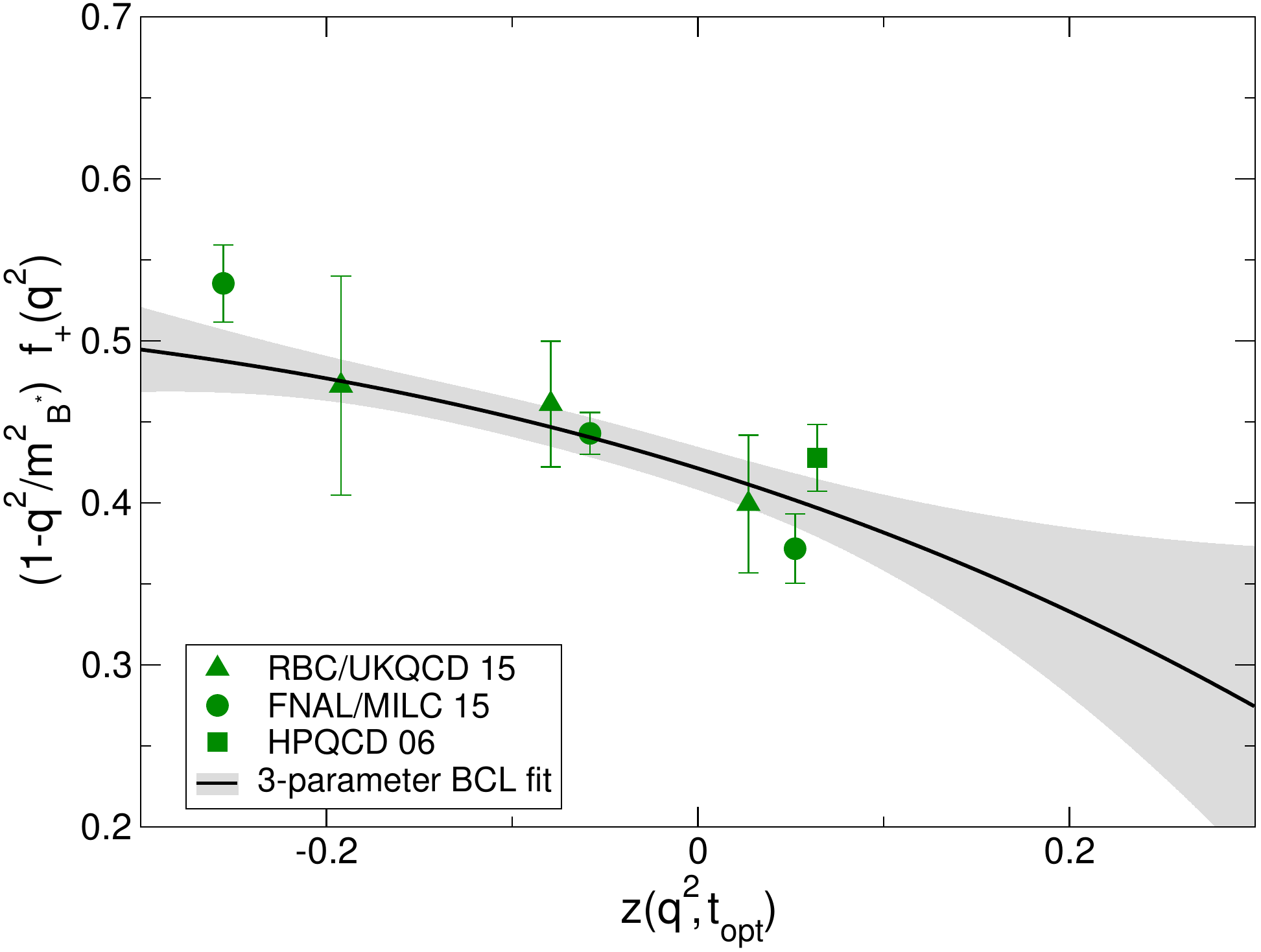}}
\begin{minipage}{0.45\textwidth}
\includegraphics[width=\textwidth]{./figs/fp_Bpi_latt-eps-converted-to.pdf}
\end{minipage}
\begin{minipage}{0.45\textwidth}
   \llap{\makebox[\wd1][l]{\raisebox{54mm}{
   \hspace{6mm}\includegraphics[width=0.15\textwidth]{./figs/FLAG_Logo-eps-converted-to.pdf}
   }}}
\end{minipage}
\vspace{-3mm}
\caption{FLAG-3 BCL fit (grey band) for the form factor $f_+(q^2)$
for $B \to \pi\ell\nu$, plotted versus $z(q^2)$.
(See~\cite{FLAG3} for a complete list of references and a discussion of dataset
and fit details.)}
\label{fig:Btopi}
\end{center}
\end{figure}

The process $B\to\pi\ell\nu_\ell$ for light final leptons is well-controlled experimentally;
the state-of-the-art experimental dataset comprises various BaBar and Belle measurements,
both untagged and with different tagging methods~\cite{bpi}.
The HFAG average for the total branching fraction has a 3\% error~\cite{Amhis:2014hma}, and the measured form-factor
shapes show good agreement. Belle II projections foresee a precision improvement by a factor of~4
with the full $50~{\rm ab}^{-1}$ dataset, measured in terms of the error on $|V_{ub}|$~\cite{BelleIIproj}.
Meanwhile, the process $B_s\to K\ell\nu_\ell$ has not been measured yet, though it is expected
to be within reach of the upcoming generation of $B$-physics results.

Until recently, only two $\NF=2+1$ results for $B\to\pi\ell\nu$ were available, by FNAL/MILC~\cite{Bailey:2008wp}
(addressing $f_+$ only) and HPQCD~\cite{Dalgic:2006dt} (with both $f_+$ and $f_0$). The last two years saw however
remarkable progress: FNAL/MILC significantly improved their previous determination, and produced
results also for $f_0$~\cite{Lattice:2015tia}; while RBC/UKQCD added their own independent determination of both
form factors~\cite{Flynn:2015mha}. The latter paper also addresses $B_s\to K\ell\nu$, which, together with a
previous publication by HPQCD~\cite{Bouchard:2014ypa}, brings a SM prediction for the rate of this process.
It is interesting to note that there is good consistency between all the computations, save
for the scalar form factor in $B\to\pi\ell\nu$: in that case, the pre-existing HPQCD determination
exhibits a discrepancy with the new FNAL/MILC result (which is in turn consistent with
the however less-precise RBC/UKQCD determination) at more than
three standard deviations.

\begin{figure}[t!]
\begin{center}
\setbox1=\hbox{\includegraphics[width=0.45\textwidth]{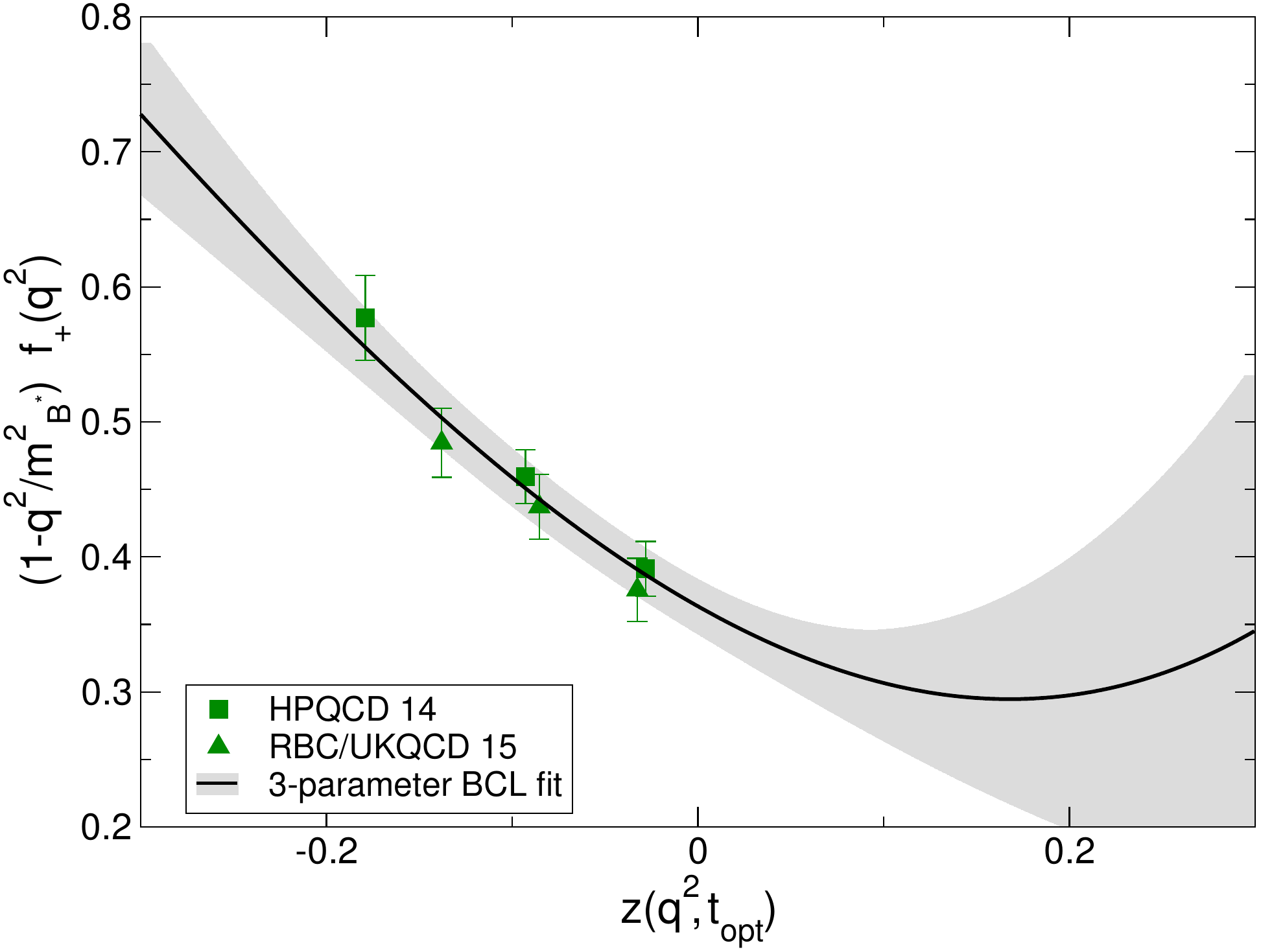}}
\setbox2=\hbox{\includegraphics[width=0.45\textwidth]{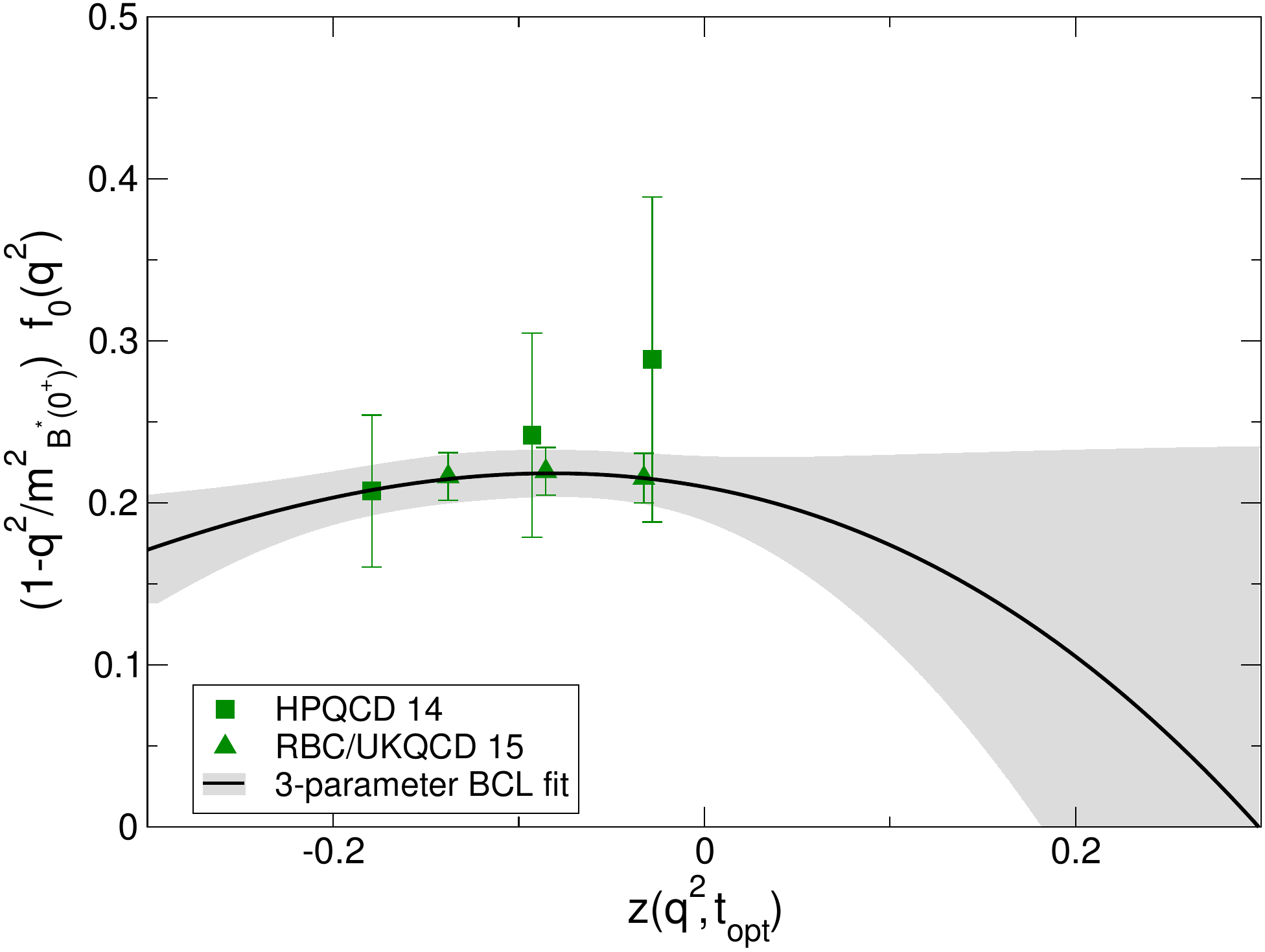}}
\begin{minipage}{0.45\textwidth}
\includegraphics[width=\textwidth]{./figs/fp_BsK_latt-eps-converted-to.pdf}
\end{minipage}
\begin{minipage}{0.45\textwidth}
\includegraphics[width=\textwidth]{./figs/f0_BsK_latt-eps-converted-to.pdf}
\end{minipage}
\begin{minipage}{0.45\textwidth}
   \llap{\makebox[\wd1][l]{\raisebox{55mm}{
   \hspace{-66mm}\includegraphics[width=0.15\textwidth]{./figs/FLAG_Logo-eps-converted-to.pdf}
   }}}
\end{minipage}
\begin{minipage}{0.45\textwidth}
   \llap{\makebox[\wd2][l]{\raisebox{55mm}{
   \hspace{4mm}\includegraphics[width=0.15\textwidth]{./figs/FLAG_Logo-eps-converted-to.pdf}
   }}}
\end{minipage}
\vspace{-3mm}
\caption{FLAG-3 BCL fits (grey band) for form factors $f_+(q^2)$
(left) and $f_0(q^2)$ (right) for $B_s \to K\ell\nu$, plotted versus $z(q^2)$.
(See~\cite{FLAG3} for a complete list of references and a discussion of dataset
and fit details.)}
\label{fig:BstoK}
\end{center}
\end{figure}

\refig{fig:Btopi} shows the FLAG-3 averaged vector form factor for $B\to\pi\ell\nu$, using
a BCL ansatz for the $q^2$ dependence (see \res{sec:percent}); this result, strongly dominated by the new FNAL/MILC
determination, allows for a significant improvement in the exclusive determination of $|V_{ub}|$
(cf.~\res{sec:CKM}). Due to the discrepancies for $f_0$ mentioned above, FLAG-3 has instead
not provided an average for the latter. In the case of $B_s\to K\ell\nu$, the good consistency
of the results by HPQCD and RBC/UKQCD has instead allowed to average both form factors,
illustrated in \refig{fig:BstoK}. In this case the precision for the vector form factor is comparable
between the two computations, while the RBC/UKQCD determination of $f_0$ is much more
precise than HPQCD's.

\subsection{$\Lambda_b\to \Lambda_c\ell\nu$ and $\Lambda_b\to p\ell\nu$ decays}

\noindent
A very interesting new development in LQCD computations for heavy quark physics is
the study of semileptonic decays of the $\Lambda_b$ baryon, with first unquenched results
provided in a work by Detmold, Lehner and Meinel~\cite{Detmold:2015aaa}. The computation is
based on RBC/UKQCD $N_f=2+1$ DWF ensembles,
and treats the $b$ and $c$ quarks within the Columbia RHQ approach. The importance of this
result is that, together with a recent analysis by LHCb of the ratio of
decay rates $\Gamma(\Lambda_b\to p\ell\nu)/\Gamma(\Lambda_b\to \Lambda_c\ell\nu)$~\cite{Aaij:2015bfa},
it allows for an exclusive determination of the ratio $|V_{ub}|/|V_{cb}|$ largely
independent from the outcome of different exclusive channels, thus contributing
a very interesting piece of information to the existing tensions in the determination
of third-column CKM matrix elements (cf.~\res{sec:CKM}).

The amplitudes of the decays $\Lambda_b\to p\ell\nu$ and $\Lambda_b\to \Lambda_c\ell\nu$
receive contributions from both the vector and the axial components of the current
in the matrix elements $\langle p|\bar q\gamma^\mu(\mathbf{1}-\gamma_5)b|\Lambda_b\rangle$
and $\langle \Lambda_c|\bar q\gamma^\mu(\mathbf{1}-\gamma_5)b|\Lambda_b\rangle$,
and can be parameterized in terms of six different form factors~\cite{Feldmann:2011xf} ---
three in the parity-even sector and three in the parity-odd sector.
All of them provide parametrically comparable contributions.
Detmold et al. obtain results for all these form factors from suitable three-point functions,
and fit them to a modified $z$-expansion ansatz (cf.~\res{sec:percent}) that combines the $q^2$ dependence
with the chiral and continuum extrapolations. The relevant systematics is obviously
very different with respect to the computations discussed above, since here baryonic
channels are involved. The main results of the paper are
the predictions for the individual form factor shapes and for the integrated
rates in the interval of momentum transfer employed in the LHCb measurement.
Prediction for the total rates in all possible
lepton channels, as well as for ratios similar to $R(D)$ (cf. above) between the $\tau$
and light lepton channels are also available.

\subsection{Rare decays}

\noindent
LQCD input is available for some exclusive semileptonic
decay channels involving neutral-current $b\to q$ transitions at the
quark level, where $q=d,s$. Being forbidden at tree level in the SM, these processes
allow for stringent tests of new physics; relevant examples
are $B\to K^*\gamma$, $B\to K^{(*)}\ell^+\ell^-$, or $B\to\pi\ell^+\ell^-$.

The corresponding SM effective
weak Hamiltonian is considerably more complicated than the one for
tree-level processes: after neglecting top quark
effects, as many as ten dimension-six operators formed by the product
of two hadronic currents or one hadronic and one leptonic current
appear.\footnote{See, e.g., \cite{Antonelli:2009ws} and references
therein.}  Three of the latter, coming from penguin and box diagrams,
dominate at short distances; supplementing this with a combination of high-energy
OPE arguments and results from Soft Collinear Effective Theory
at intermediate energies, it is possible to argue that their contributions
are still dominant when long-distance physics is also taken into account.
Within this approximation, the dominant long-distance contribution thus
consists of matrix elements of current operators (vector,
tensor, and axial-vector) between one-hadron states, which in turn can
be parameterized in terms of a number of form factors~\cite{Liu:2009dj}.
On top of the aforementioned approximations, the
lattice computation of the relevant form factors in channels with a
vector meson in the final state faces extra challenges on top of those
already present in the case of a pseudoscalar meson: the state
is unstable, and the extraction of the relevant matrix element from
correlation functions is significantly more complicated; and $\chi$PT
cannot be used as a guide to extrapolate results at unphysically heavy
pion masses to the chiral limit. While the field theory procedures to take
resonance effects into account are available~\cite{fvscat},
they have not yet
been implemented in the existing preliminary computations, which therefore
suffer from an essentially uncontrolled source of systematic uncertainty.\footnote{This is
a non-negligible effect e.g. in $B\to K^*$ transitions, given the $K^*$ width.}

In decays to pseudoscalar mesons, there are results
for the vector, scalar, and tensor form factors for $B_s\to K\ell^+\ell^-$ decays
by HPQCD~\cite{Bouchard:2013pna} and FNAL/MILC~\cite{Bailey:2015dka}, the
latter paper also providing results for $B\to\pi\ell^+\ell^-$.
Concerning channels with vector mesons in the final state, Horgan {\it et al.}
have obtained the seven form factors relevant for $B \to K^* \ell^+
\ell^-$ (as well as those for $B_s \to \phi\, \ell^+ \ell^-$) in
\cite{Horgan:2013hoa} using NRQCD $b$ quarks and asqtad staggered
light quarks. Finally,
ongoing work on $B\to K^*\ell^+\ell^-$ and $B_s\to \phi\ell^+\ell^-$
by RBC/UKQCD, including first results, has recently been reported in~\cite{Flynn:2015ynk}.

\section{CKM matrix elements}
\label{sec:CKM}

\noindent
The LQCD-determined decay constants and form factors discussed above can be combined
with the relevant experimental results to obtain exclusive values for the second- and third-row CKM matrix
elements $|V_{cd}|$, $|V_{cs}|$, $|V_{ub}|$, and $|V_{cb}|$. In the case of $|V_{cq}|$,
determinations are possible with results for $\NF=2,~2+1,~2+1+1$; they mostly come from
leptonic decays, with just one $\NF=2+1$ semileptonic determination based on HPQCD's form factors.
For $|V_{cb}|$ only semileptonic determinations based on $\NF=2+1$ (two channels)
and $\NF=2$ (one channel) computations are possible. Finally, for $|V_{ub}|$ both semileptonic
($\NF=2+1$) and leptonic ($\NF=2,~2+1,~2+1+1$) determinations are possible;
however, the latter come from the still poorly understood $B\to\tau\nu_\tau$ measurements,
and have much larger errors than the semileptonic determination from $B\to\pi\ell\nu$.
In the remainder of this section we will briefly summarise the FLAG-3 updated
determination of these CKM matrix elements, based on the averages discussed in previous sections.

\begin{figure}[t!]
\begin{center}
\begin{minipage}{0.45\textwidth}
\includegraphics[width=\textwidth]{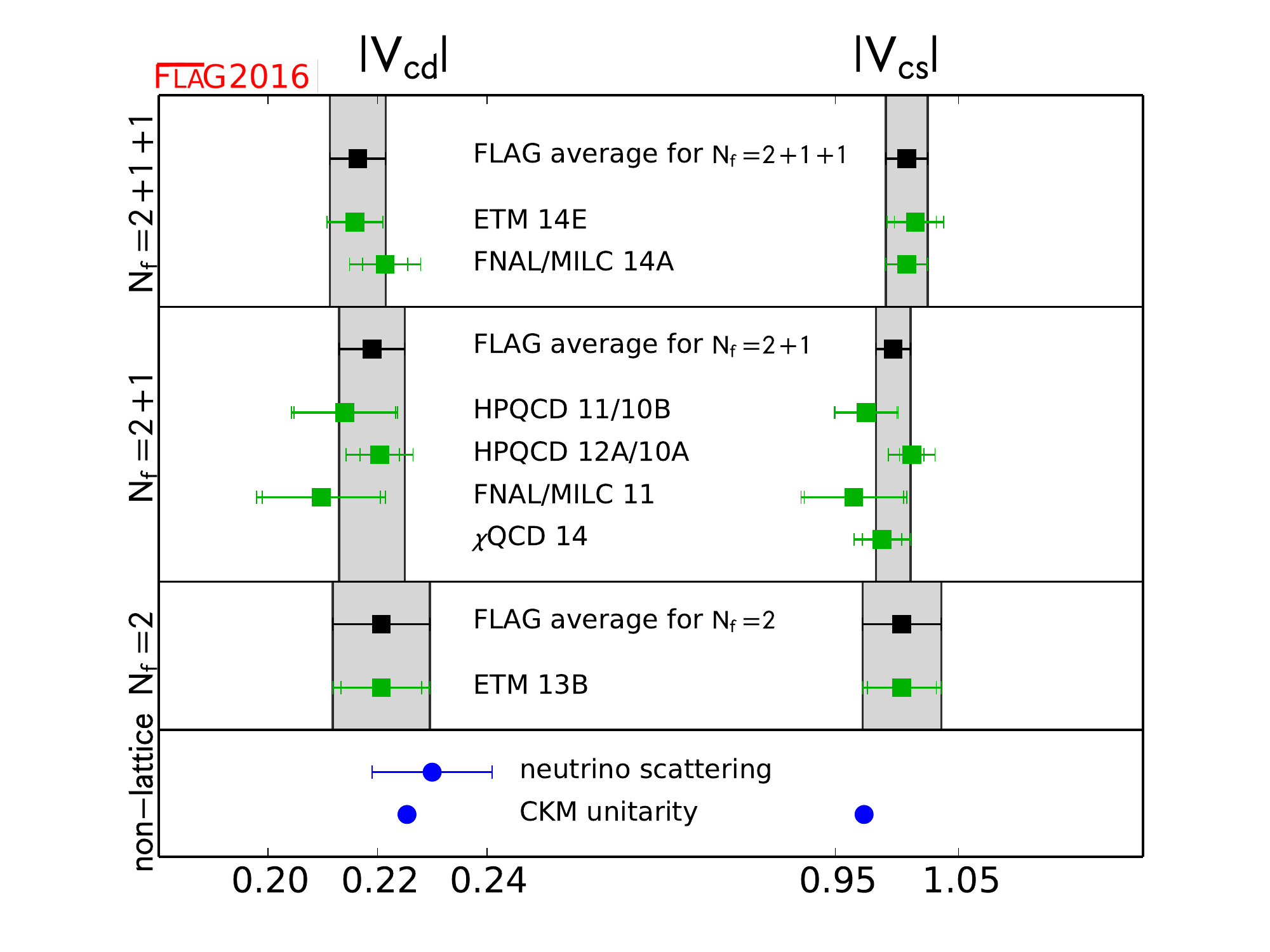}
\end{minipage}
\vspace{-2mm}
\caption{Determinations of $|V_{cd}|$ and $|V_{cs}|$ from values of charm decay constants
and form factors entering FLAG-3 averages.
(See~\cite{FLAG3} for a complete list of references.)}
\label{fig:VcdVcs}
\end{center}
\end{figure}

In order to determine $|V_{cd}|$ and $|V_{cs}|$, FLAG-3 combine their averages
with PDG averages for $f_D|V_{cd}|$ and $f_{D_s}|V_{cs}|$~\cite{Agashe:2014kda}, and HFAG
averages for $f^{D\to\pi}_+(0)|V_{cd}|$ and $f^{D\to K}_+(0)|V_{cs}|$~\cite{Amhis:2014hma}.
The result for all computations contributing to averages is illustrated
in~\refig{fig:VcdVcs}; the resulting average values are provided in~\ret{tab:VcdVcs}.
There is good consistency among all the determinations --- in particular, no $\NF$
dependence is apparent. The averages are consistent with $|V_{cd}|^2+|V_{cs}|^2+|V_{cb}|^2=1$
within at most two standard deviations; the level of precision makes this result
independent of the value employed for $|V_{cb}|$.

\begin{figure}[t!]
\begin{center}
\setbox1=\hbox{\includegraphics[width=0.45\textwidth]{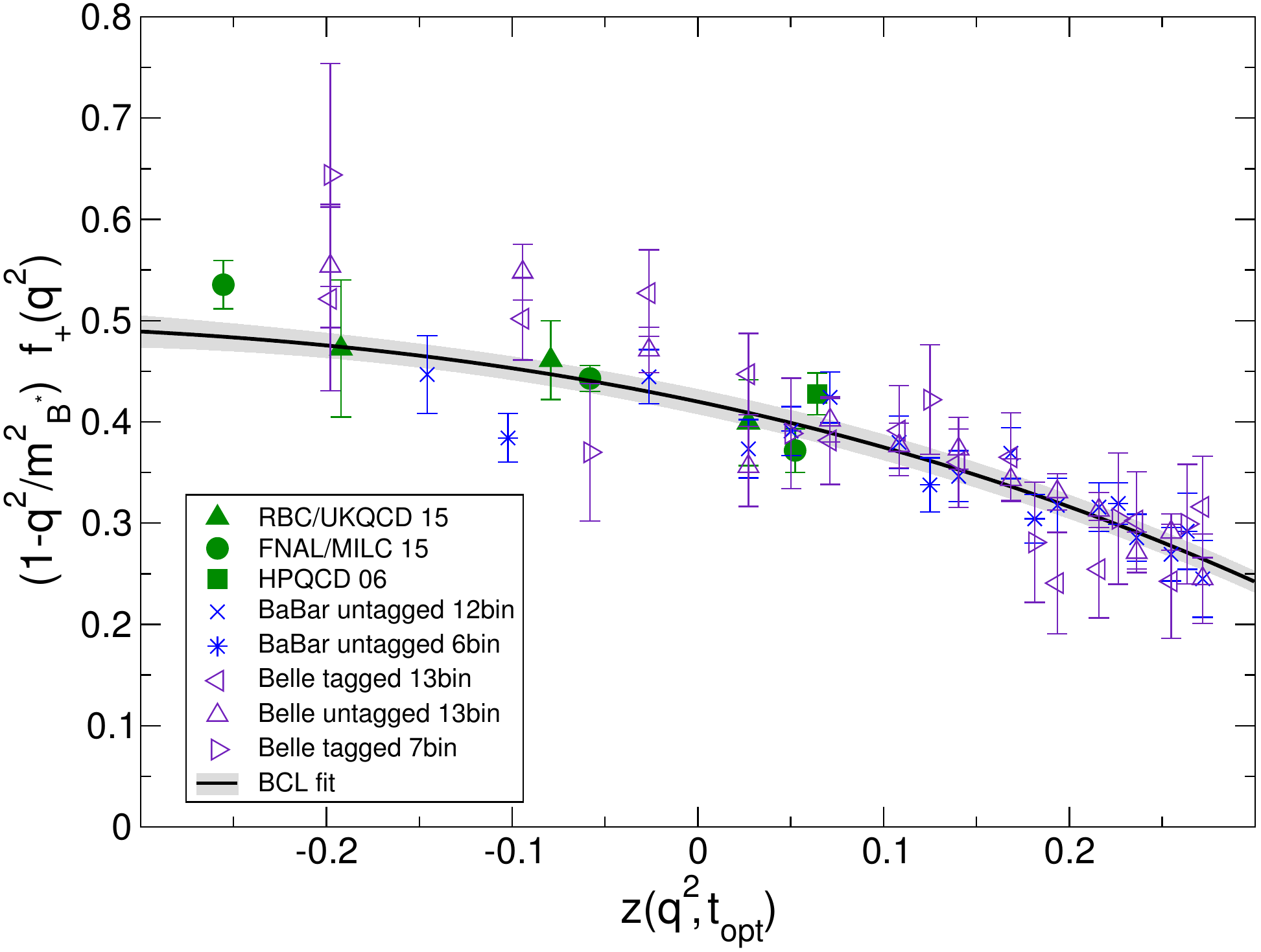}}
\setbox2=\hbox{\includegraphics[width=0.45\textwidth]{./figs/fp_BD_latt-eps-converted-to.pdf}}
\begin{minipage}{0.45\textwidth}
\includegraphics[width=\textwidth]{./figs/fp_Bpi_latt+exp-eps-converted-to.pdf}
\end{minipage}
\begin{minipage}{0.45\textwidth}
\includegraphics[width=\textwidth]{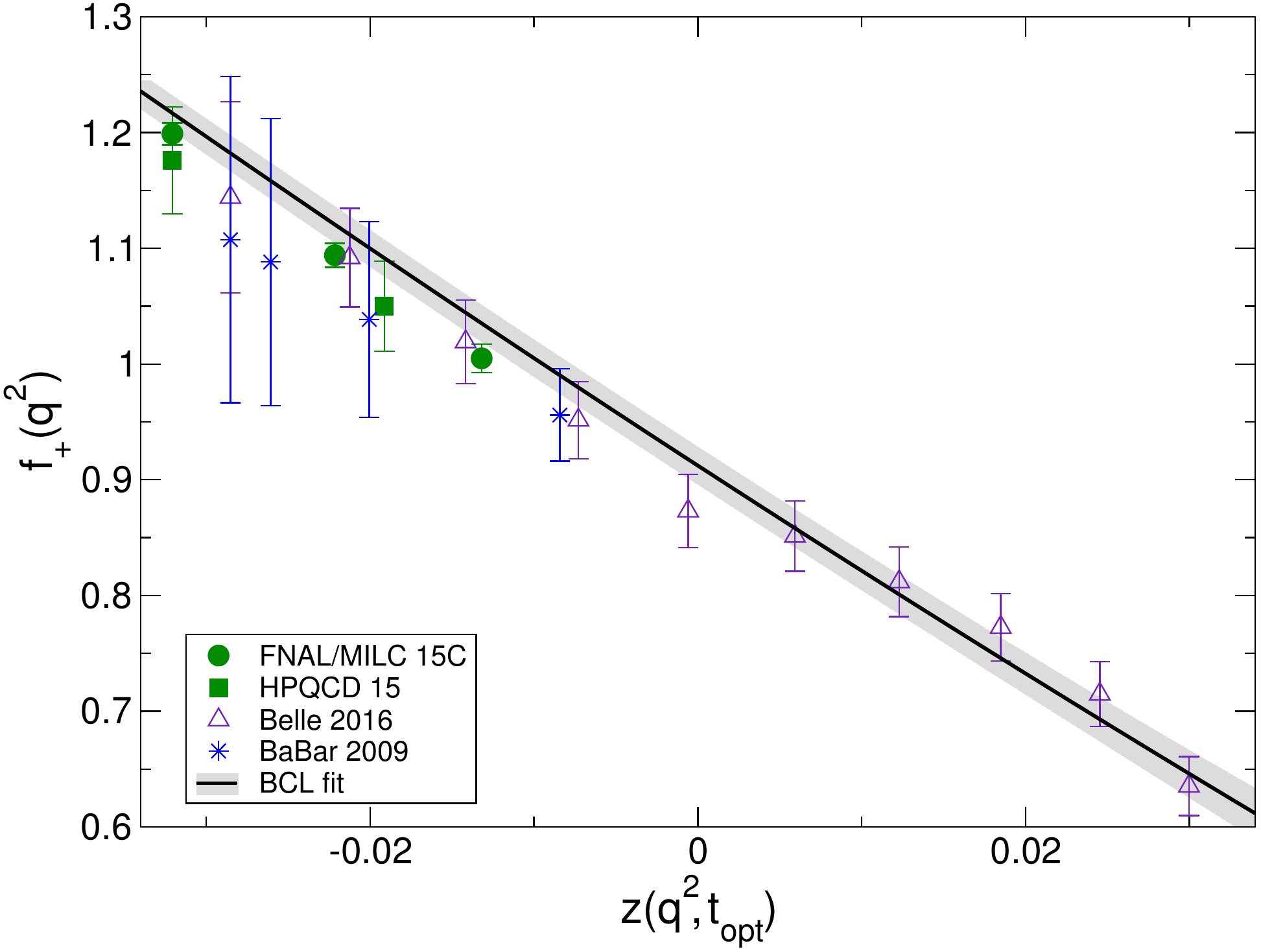}
\end{minipage}
\begin{minipage}{0.45\textwidth}
   \llap{\makebox[\wd1][l]{\raisebox{55mm}{
   \hspace{-66mm}\includegraphics[width=0.15\textwidth]{./figs/FLAG_Logo-eps-converted-to.pdf}
   }}}
\end{minipage}
\begin{minipage}{0.45\textwidth}
   \llap{\makebox[\wd2][l]{\raisebox{55mm}{
   \hspace{4mm}\includegraphics[width=0.15\textwidth]{./figs/FLAG_Logo-eps-converted-to.pdf}
   }}}
\end{minipage}
\vspace{-3mm}
\caption{FLAG-3 BCL joint fits (grey band) to lattice and experimental
form-factor data for $B\to\pi\ell\nu$ (left) and $B\to D\ell\nu$ (right), plotted versus $z(q^2)$.
(See~\cite{FLAG3} for a complete list of references and a discussion of dataset
and fit details.)}
\label{fig:CKMfits}
\end{center}
\end{figure}

\begin{figure}[t!]
\begin{center}
\begin{minipage}[t]{0.45\textwidth}
\vspace{-0pt}
\includegraphics[width=\textwidth]{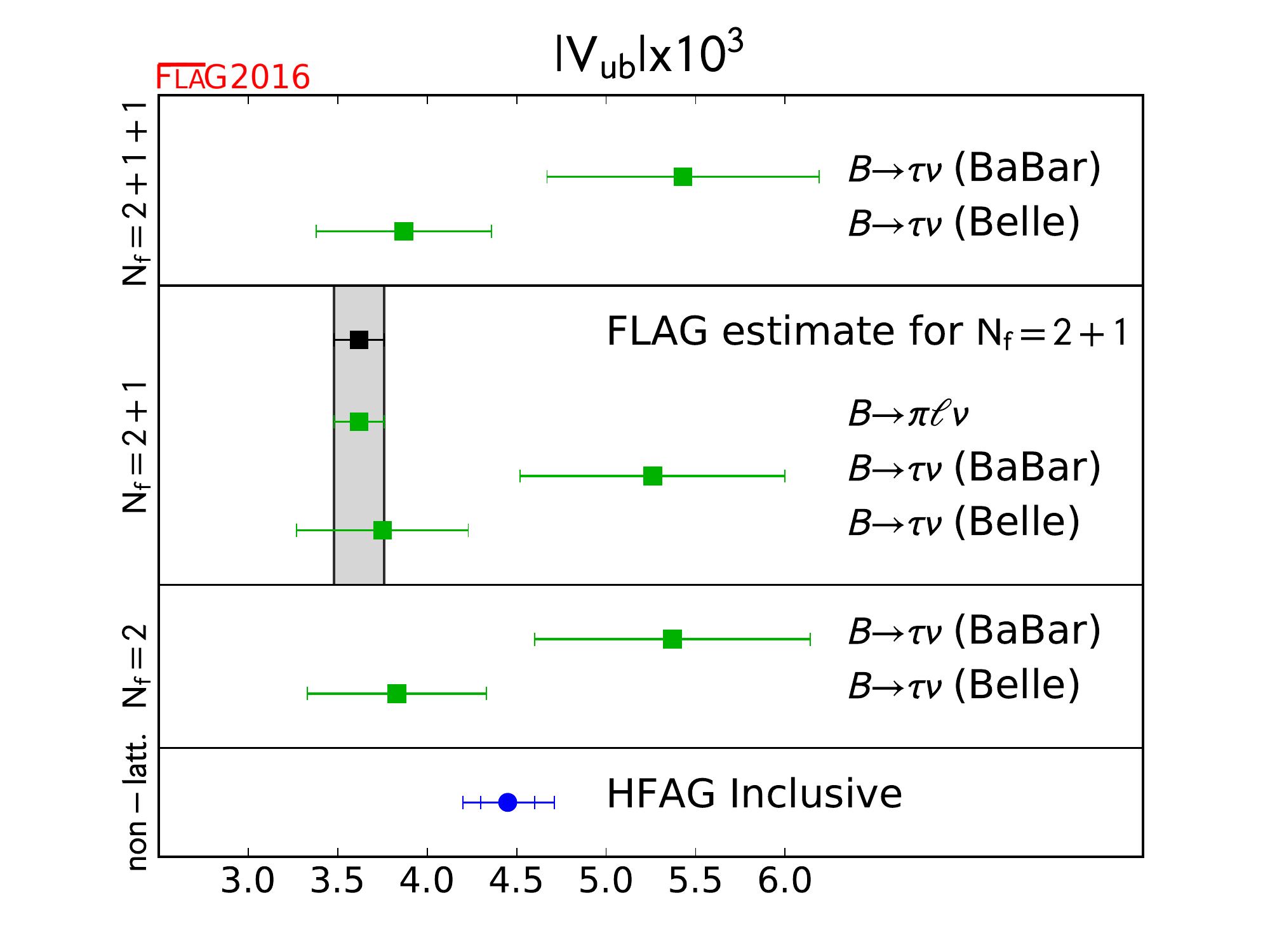}
\end{minipage}
\hspace{10mm}
\begin{minipage}[t]{0.45\textwidth}
\vspace{-0pt}
\includegraphics[width=\textwidth]{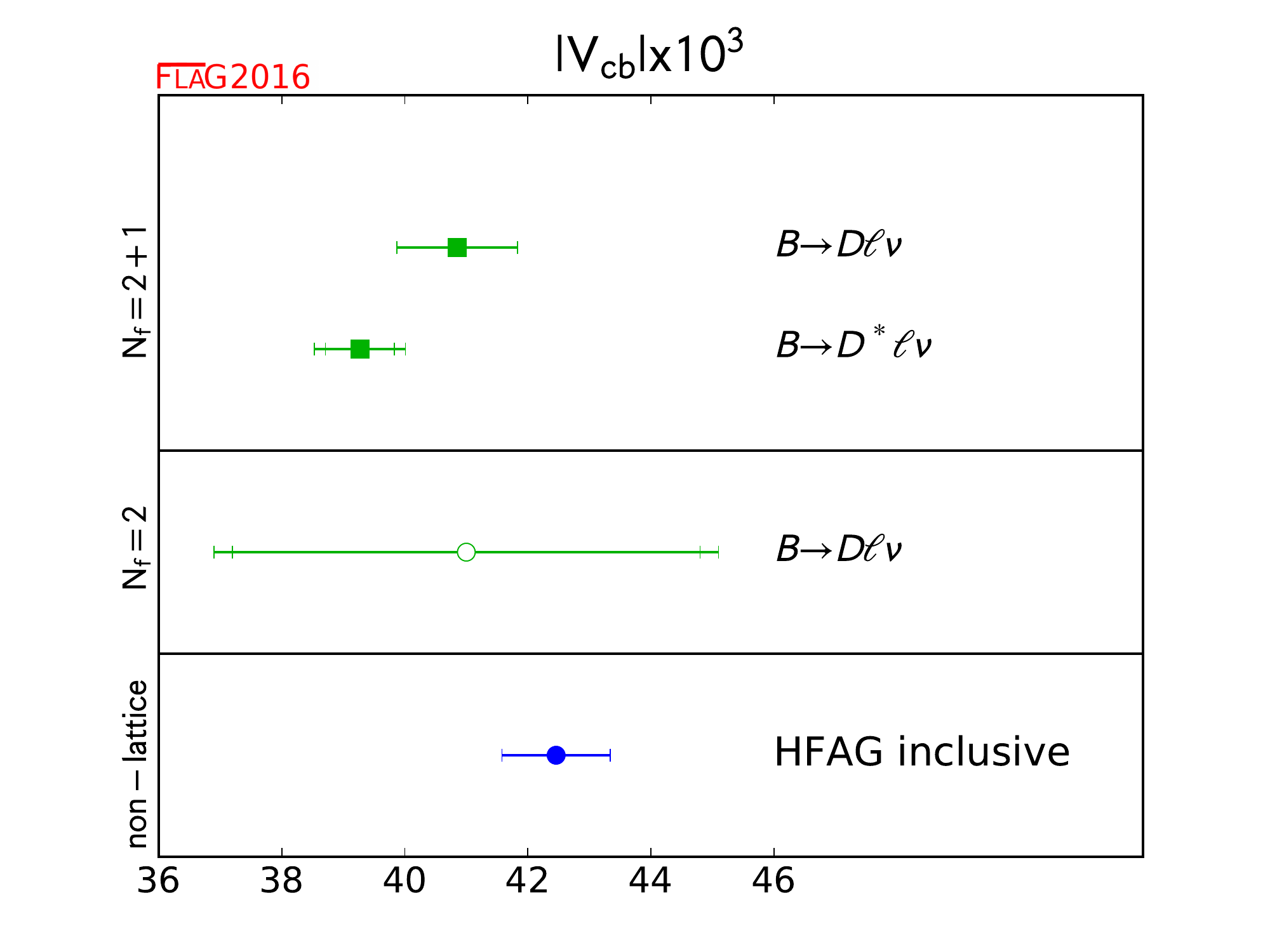}
\end{minipage}
\end{center}
\vspace{-7mm}
\caption{FLAG-3 summary plots for the determination of $|V_{ub}|$ and $|V_{cb}|$ (see~\cite{FLAG3} for a complete list of references).}
\label{fig:VubVcb}
\end{figure}

In the case of $|V_{ub}|$ and $|V_{cb}|$, accurate determinations based on $B\to\pi\ell\nu$
and $B\to D\ell\nu$, respectively, can be obtained from simultaneous fits to the lattice
vector form factors and state-of-the-art experimental data as a funcion of $q^2$, using a BCL
ansatz (see \res{sec:percent}) in which the CKM is left as a fitted relative normalisation. The outcome of this
exercise, using as input the FLAG-3 averages for the form factors, is shown in \refig{fig:CKMfits}.
The resulting CKM values are shown in~\refig{fig:VubVcb}, together with leptonic determinations
of $|V_{ub}|$ and the $|V_{cb}|$ determinations based on $B\to D^*\ell\nu$, as well as with
inclusive determinations. The well-known tension between inclusive and exclusive values
is still present with the latest generation of lattice results. FLAG-3 has decided
not to quote a value for $|V_{ub}|/|V_{cb}|$ based on $\Lambda_b$ decays, since
the lattice results for the latter do not meet some of the FLAG-3 requirements
to enter averages or estimates. A summary of FLAG-3 quoted CKM values is provided in~\ret{tab:VubVcb}.

\begin{table}[t!]
\begin{center}
\begin{tabular}{lcll}
\Hline
\multicolumn{1}{c}{$\NF$} & from & \multicolumn{1}{c}{$|V_{cd}|$} & \multicolumn{1}{c}{$|V_{cs}|$} \\
\hline
$2+1+1$ & $f_D$ and $f_{D_s}$ & 0.2164(51) & 1.008(17) \\
$2+1$   & $f_D$ and $f_{D_s}$ & 0.2195(61) & 1.004(18) \\
$2$     & $f_D$ and $f_{D_s}$ & 0.2207(89) & 1.004(32) \\
\hline
$2+1$   & $D\to\pi\ell\nu$ and $D\to K\ell\nu$ & 0.2140(97) & 0.975(26) \\
\hline
$2+1$   & L+SL average & 0.2190(60) & 0.997(14) \\
\Hline
\end{tabular}
\end{center}
\vspace{-5mm}
\caption{FLAG-3 determinations of $|V_{cd}|$ and $|V_{cs}|$. ``L+SL'' refers to the (correlated) average between the leptonic and semileptonic determinations.}
\label{tab:VcdVcs}
\end{table}

\section{The percent precision target}
\label{sec:percent}

\noindent
As shown above, the precision level already attained for several observables
of interest in heavy flavour physics on the lattice is at the few percent level.
Exclusive CKM determinations bear errors generally better than $4\%$,
and the theory and experimental uncertainties are generally comparable.
The leap on experimental precision expected from the current and upcoming generations of experiments
will thus pose a stiff challenge to LQCD computations, which are expected
in many cases to reach precisions in the $1$--$2\%$ ballpark. This is indeed
already commonplace in computations in the pion and kaon sector; therefore,
it is crucial to focus on the specific systematic uncertainties appearing
in heavy flavour computations.
One common aspect with light hadron physics is the need to consistently incorporate
electromagnetic corrections, which are now the subject of intense work ---
see, e.g.,~\cite{qed};
or higher-order OPE contributions, which are already being considered in kaon observables~\cite{kpipild}.
Issues related to the difficulty to treat the $b$ quark within practical lattice regularisations,
the correct treatment of resonances, or the use of perturbation theory, have already
been briefly touched upon above.

The computation of semileptonic decay amplitudes has its own share of specific
issues to be dealt with, like the effect of contributions from excited states~\cite{Bahr:2016ayy},
or the lack of an adequate chiral perturbation theory description in channels
where final-state pseudoscalar mesons can have energies much larger than the chiral cutoff
(e.g.~$B\to\pi\ell\nu$).\footnote{See e.g. the discussion in~\cite{Lattice:2015tia}, and the related works~\cite{hpchipt}.}
One further key source of systematic uncertainties, which is now becoming crucial for several
decay channels, is the description of the momentum transfer dependence of form factors;
let us now conclude with a brief survey of this issue.

The benchmark channel where this systematics has long been studied is $B\to\pi\ell\nu$,
since in this channel the kinematically allowed region in $q^2$ is broad enough so as
to make a description in terms of a form factor at fixed $q^2$ impractical.
Ans\"atze for the $q^2$ dependence are based on the generic observation that
all form factors are analytic functions on the complex $q^2$ plane outside physical poles
and inelastic threshold branch points; and this process is particularly simple,
since the only resonance pole below the $B\pi$ production region is the (narrow) $B^*$,
which is furthermore close to the threshold. Simple ansatz choices can thus be constructed
using the idea of pole dominance; in particular, the Be\'cirevi\'c-Kaidalov~\cite{Becirevic:1999kt}
and Ball-Zwicky~\cite{Ball:2004ye} descriptions have been widely used.
A more systematic approach exploits the positivity and analyticity properties of
two-point functions of the vector current to obtain optimal parametrisations of form
factors~\cite{zpar}.
The general form of these so-called $z$-parametrisations is
\begin{gather}
f(q^2) = \frac{1}{B(q^2)\phi(q^2,t_0)}\,\sum_{n=0}^\infty a_n(t_0)\,z(q^2,t_0)^n\,, \qquad
z(q^2,t_0) = \frac{\sqrt{t_+-q^2}-\sqrt{t_+-t_0}}{\sqrt{t_+-q^2}+\sqrt{t_+-t_0}}\,,
\end{gather}
where the latter kinematical variable amounts to a conformal transformation of the
complex $q^2$ plane, dependent on an essentially arbitrary parameter $t_0$, that maps the
semileptonic region into a disc; $B(q^2)$ is a Blaschke factor that contains sub-threshold poles;
and the outer function $\phi$ is some smooth positive function of $q^2$. The crucial property
of this series expansion is that the coefficients $a_n$ satisfy the unitarity bound
\begin{gather}
\sum_{n=0}^\infty a_n^2 = \frac{1}{2\pi i}\oint \frac{dz}{z}\,|B(z)\phi(z)f(z)|^2
\end{gather}
(where $\oint$ is taken over the image disc boundary in the $z$ plane).
With a judicious choice for $\phi$ and $t_0$, this translates into a strong constraint,
that allows to describe the form factor in terms of few free parameters, {\em and}
a solid bound on the systematic uncertainty due to the series truncation.
The simplest choice for the $B\to\pi$ vector form factor, dubbed BCL after the authors of~\cite{Bourrely:2008za},
is
\begin{gather}
\label{eq:bcl}
f_+(q^2)=\frac{1}{1-q^2/m_{B^*}^2}\,\sum_{n=0}^N a_n(t_0) z(q^2,t_0)^n\,, \qquad
t_0 = (m_B+m_\pi)(\sqrt{m_B}-\sqrt{m_\pi})^2\,,
\end{gather}
with the additional constraint $a_N=-\,\frac{(-1)^N}{N}\,\sum_{n=0}^{N-1}(-1)^n\,n\,a_n$
on the highest-order coefficient of the truncated series, that imposes the correct
asymptotic behaviour at threshold. \req{eq:bcl} is the FLAG-recommended parametrisation
of choice for form factors.

\begin{table}[t!]
\begin{center}
\begin{tabular}{l|cl|cl}
\Hline
\multicolumn{1}{c|}{$\NF$} & from & \multicolumn{1}{c|}{$|V_{ub}|$} & from & \multicolumn{1}{c}{$|V_{cb}|$} \\
\hline
\multirow{2}{*}{$2+1$} & $B\to\pi\ell\nu$ & $3.62(14)\times 10^{-3}$ & $B\to D\ell\nu$ & $40.85(98)\times 10^{-3}$ \\
  & & & $B\to D^*\ell\nu$ & $39.27(74)\times 10^{-3}$ \\
\hline
$2$     & & & $B\to D\ell\nu$ & $41.0(4.1)\times 10^{-3}$ \\
\Hline
\end{tabular}
\end{center}
\vspace{-5mm}
\caption{FLAG-3 determinations of $|V_{ub}|$ and $|V_{cb}|$.}
\label{tab:VubVcb}
\end{table}

The discussion above largely extends to the scalar form factor in $B\to\pi\ell\nu$ decays,
as well as to form factors for other semileptonic transitions;
a general discussion can be found, e.g., in~\cite{Hill:2006ub}. The form factors for a generic $H \to L$
transition will display a cut starting at the production threshold $t_+$, and the optimal
value of $t_0$ required in $z$-parameterizations is $t_0=t_+(1-\sqrt{1-t_-/t_+})$
(where $t_\pm=(m_H\pm m_L)^2$).
For unitarity bounds to apply, the Blaschke factor has to include all sub-threshold
poles with the quantum numbers of the hadronic current --- e.g., vector (resp. scalar)
resonances with $b\bar u$ quark content for the $B\to\pi$, $B_s\to K$ vector (resp. scalar) form factors;
and idem $b \bar c$ quark content for $B\to D$.\footnote{A more complicated analytic structure
may arise in other cases, such as channels with vector mesons in the final state.}
Thus, as emphasized above, the control over systematic uncertainties brought in by using
$z$-parametrisations strongly depends on implementation details.
This has practical consequences, in particular, when the resonance spectrum
in a given channel is not sufficiently well-known, or resonances are very
close to thresholds, and may go across them as quark masses are changed. Caveats may also
apply for channels where resonances with a non-negligible width appear.
A further issue is whether $t_+=(m_H+m_L)^2$ is the proper choice for the start of the cut in cases such as $B_s\to K\ell\nu$ and $B\to D\ell\nu$, where there are lighter two-particle states that project on the current ($B\pi$ and $B_c\pi$, respectively). In any such
situation, it is not clear a priori that a given $z$-parametrisation will
satisfy strict bounds, as has been seen, e.g., in determinations of the proton charge radius
from electron-proton scattering~\cite{paz}.
One particular case where several of these issues may be at play regards
the use of so-called modified $z$-expansions, pioneered by HPQCD, that
take into account simultaneously the $q^2$, light quark masses, and
lattice spacing dependence of form factors. Because the modified $z$-expansion is not
derived from an underlying effective field theory, it is unclear
how systematic uncertainties can be quantified within this approach
--- in particular, the applicability of unitarity bounds has to be examined carefully.

The rapid growth in the number of lattice and experimental results for semileptonic
decays where an accurate description of the $q^2$ dependence is needed warrants an
in-depth general study of how these ideas can be optimally applied to as many channels
as possible. One particularly interesting case, from the methodological point of view,
is the recent extraction of accurate $D\to\pi$ and $D\to K$ form factors by BESIII~\cite{Ma:2015tja},
where the dominant source of experimental uncertainty in view of CKM determinations
comes from the description of the $q^2$ dependence. Since the relevant vector resonances
are extremely close to the inelastic threshold, these channels constitute an excellent
laboratory to deepen into a number of the issues pointed out above.

\section{Conclusions and outlook}
\label{sec:conclusions}

\noindent
There is rapid progress in LQCD relevant for heavy quark physics --- especially so,
during the last two years, in addressing semileptonic $b$ decays. For several
observables of interest (e.g. those involved in the determination of CKM matrix elements)
the theory uncertainty is now comparable to, or smaller than, the experimental one.
There is also a fertile interaction between lattice and experimental collaborations,
which allows to better focus efforts on both sides; the recent study of $\Lambda_b$
decays is an excellent success story in this respect.

The upcoming era of experimental results, led by LHCb, BESIII and, especially, Belle II,
will however pose a strong challenge to the accuracy of lattice methods. In particular,
the systematic uncertainties related to the treatment of $b$ quarks will take a central
role, and it will become crucial to have as many cross-checks as possible among different
procedures. Also, many small effects which have been up to now neglected or not fully
addressed --- from electromagnetic corrections to various field-theory aspects
(use of perturbation theory, chiral extrapolations, $q^2$ dependence of form factors, resonance effects) ---
will become relevant.
Incorporating state-of-the-art ensembles already used for pion and kaon
physics to new heavy-flavour studies will play a role in decreasing various uncertainties, too.

Ultimately, however, it is of the utmost importance to diminish our reliance on effective
theories by being able to directly simulate physical $b$ and light quarks simultaneously.
Important steps in that direction are being taken~\cite{obc,Mages:2015scv,Endres:2015yca},
and should have a key role
in planning for new $\NF=2+1+1(+1)$ simulations in the near future.

\section*{Acknowledgements}

\noindent
Support from the EU PITN-GA-2009-238353 (STRONGnet), MCINN grants FPA2012-31686 and FPA2012-31880, and
MINECO’s “Centro de Excelencia Severo Ochoa” Programme under grant SEV-2012-0249,
is gratefully acknowledged.
In the preparation of my talk I benefited from discussions with, and input from,
C.T.H. Davies, C. DeTar, P. Dimopoulos, D. Du, G. Herdo\'{\i}za,
T. Ishikawa, A. Kronfeld, P. Lami, K. Nakayama, R. Sommer,
J.T. Tsang, C. Urbach, R. Van de Water, and Y. Yang.
I am indebted to my FLAG colleagues for the huge amount of work put in producing
the review, and for many illuminating discussions. A critical reading of a first version of this writeup by P.~Fritzsch
and G.~Herdo\'{\i}za has greatly contributed towards improving it. Finally, I would
like to thank the Lattice 2015 organisers for a splendid conference, in the face
of adverse elements.
お疲れ様でした。

\end{CJK}\end{document}